%% file: emse.tex
\RequirePackage{fix-cm}
\documentclass{svjour3}                     
\smartqed  
\usepackage{graphicx}
\usepackage{booktabs}
\usepackage{rotating}
\usepackage{subcaption}
\usepackage{enumitem}
\usepackage{longtable}

\usepackage{color}
\usepackage{ifthen}
\usepackage{mdframed}
\usepackage{lscape}
\usepackage{hyperref}

\newboolean{showcomments}
\setboolean{showcomments}{true} 
\ifthenelse{\boolean{showcomments}}
    {
        \newcommand{\alvi}[1]{\textcolor{red}{{\it [Alvi says: #1]}}}
         \newcommand{\neil}[1]{\textcolor{green}{{\it [Neil says: #1]}}}

    }
    {
        \newcommand{\alvi}[1]{}
        \newcommand{\neil}[1]{}

    }

\newcommand{\replicationUrl}{\href{https://doi.org/10.5281/zenodo.4010208}{doi:10.5281/zenodo.4010208} and \href{https://doi.org/10.5281/zenodo.4885980}{https://doi.org/10.5281/zenodo.4885980}}
\newcommand{\extensionFigureUrl}{\href{https://doi.org/10.5281/zenodo.4616151}{https://doi.org/10.5281/zenodo.4616151}}
\newcommand{\SO}{Stack Overflow}

\usepackage{hyperref}

\journalname{Empirical Software Engineering}
\begin{document}

\title{Conclusion Stability for Natural Language Based Mining of Design Discussions}

\titlerunning{Design Mining Conclusion Stability}        

\author{Alvi Mahadi         \and 
        Neil A. Ernst \and
        Karan Tongay
}


\institute{A. Mahadi, N. Ernst, K Tongay \at
              Department of Computer Science \\
              University of Victoria, Canada\\
              \email{alvi.utsab@gmail.com, nernst@uvic.ca, karantongay@gmail.com}           
}

\date{Received: date / Accepted: date}

\maketitle

\begin{abstract}
Developer discussions range from in-person hallway chats to comment chains on bug reports. Being able to identify discussions that touch on software design would be helpful in documentation and refactoring software. Design mining is the application of machine learning techniques to correctly label a given discussion artifact, such as a pull request, as pertaining (or not) to design. In this paper we demonstrate a simple example of how design mining works. We then show how conclusion stability is poor on different artifact types and different projects. We show two techniques---augmentation and context specificity---that greatly improve the conclusion stability and cross-project relevance of design mining. Our new approach achieves AUC of 0.88 on within dataset classification and 0.80 on the cross-dataset classification task.
\end{abstract}


\input{sections/01-introduction}
\input{sections/02-background-related}

\input{saner_sections/03-replication}
\input{sections/deep_approach}
\input{sections/03-dataset-design-discussion}
\input{sections/04-classifying-design-discussion}
\input{sections/05-results-discussion}
\input{sections/06-discussion}
\input{sections/07-conclusion}

\begin{acknowledgements}
Thanks to the authors who have so graciously made their data available for study in true open fashion. Thanks also to the reviewers who have greatly improved this manuscript over its several iterations.
\end{acknowledgements}

\appendix
\input{sections/related_table}

\bibliographystyle{IEEEtran}
\bibliography{abbr,emse}

\end{document}

%% file: sections/01-introduction.tex
\section{Introduction}
\label{introduction}
Developer discussions play a vital role in software development. The discussions solicit the opinions of other developers and document important decisions in today's pull-based software development process \cite{Gousios2014}.  Discussions about software design, in particular, are a highly interactive process and many decisions involve considerable back and forth. These decisions greatly impact software architecture \cite{Kazman2016,woods16}. Discussions are also a rich software artifact for learning about the software itself \cite{Nazar2016}. Recent progress in research \cite{Viviani2018,Viviani2018b} suggests that developer discussions often contain rich information on the background of the design of the software as well as rationale and reflections on the design changes and choices over time. 

Such discussions are also one of the potential artifacts for newcomers to understand the architecture and design of the system \cite{Cubranic2003}. However, these discussions about design are often scattered across different places such as commit messages, pull requests, and issue tracker comments. It is impractical for anyone to go through all the thousands of threads of discussions and find out the discussion about a particular design topic. Solving this problem is the challenge of what we call \emph{design mining}, which is a branch of research based on mining software repositories.  Being able to mine design discussions would lead to a novel approach of improved documentation, enabling improved traceability of requirements, refactoring and bug fixing support,  and maintainability.      

The simplest formulation of the design mining problem is defined as classifying developer \textit{discussion} as either design or not-design\footnote{more complex formulations might identify design \emph{topic}.}. Discussions are extracted from software artifacts, including but not limited to pull requests, issues, code comments, and Q\&A interactions\footnote{and thus, design mining (to date) does not deal with more ephemeral design discussions, such as whiteboards or video conferences.}. This classification process is usually supervised: manual labelling of the data with human effort by following a coding guide, then leveraging automatic classification using machine learning models and advances in natural language processing to classify the discussions according to some specific features. 

Verification of the correctness of manual classification is achieved by meeting and agreement among the participants. Validation of automatic classification is measured by evaluating the classifiers with a manually labelled small set of data, which is  referred to as the gold set. Almost all the studies in this field have attempted to produce a best-performing and validated automated classifier \cite{Brunet2014,Shakiba2016,Viviani2018}. For example, a state of the art (SOTA) result from Viviani et al. \cite{Viviani2019} talks about a well validated model with Area Under ROC Curve (AUC) of 0.87. However, achieving conclusion stability \cite{bangash20} remains a challenge. Most studies focus on evaluating a classifier on data from a single dataset and discussion artifact. In this paper we focus on conclusion stability by developing a model with wide applicability to different design discussions which could be used with high accuracy across different projects, and artifact types. 

Automatic detection of design points can significantly reduce development time for both contributing developers as well as reviewers. It can also help to use rich design information with ease which is a struggle for newcomers to an open source project \cite{igor2014}. A recommender agent built on the detected design points can assist the core developers or maintainers to answer the question and queries from the newcomers. Because software design can be very subjective, findings from studies like this can potentially reveal several aspects of how and why design decisions diverge from ideal design patterns. Moreover, these different opinions can also be analyzed to further modernize some of the trivial design ideas. Lastly, mining and summarizing design discussions gives us a great opportunity to keep an up to date documentation with little to no manual effort and time.   

An important aspect of our view of a useful design mining classifier is its ability to work well across different projects and domains. We define the source of the data as domain. For example, if we source our data from \SO{} or Github, the domain of the data is \SO{} or Github respectively. Project is a subset of domain that is the name of the project to which the discussion belongs (ex. node.js, Rails, etc.). Cross-project/cross-domain transfer learning means training on one domain/project and transferring the knowledge to another domain/project, i.e., it shows good conclusion stability. Why is this important? In a general purpose setting, such as a bot for pull requests, contexts will change constantly. Thus a model that can perform well on (say) NodeJs pull requests and also on NASA embedded software issues would be useful for generalizability.

The second reason is that even within a company context models change. For example, new developers might change the way software design is discussed; new programming frameworks or languages are introduced; or new discussion formats---such as Github discussions---are leveraged. Finally, a research reason is that proposing models that ignore conclusion stability suggests a research approach that may only work in a particular, narrow domain of application and lack external validity. 

One of the highlights of our study is we are not restricting ourselves to specific rules to define design, such as a code book that defines what constitutes a design discussion. Instead we use publicly available data from \SO{}, generated by developers over time. This enables us to capture some unconventional notions of design in bottom-up, crowd-source fashion. By the end of this paper, we explore the following research questions:
\begin{enumerate}
    \item \emph{How can we get more labeled data to train, validate and test models of design mining?} \\
    \textbf{Approach---}We take a different approach from the previous studies to address research question 1. While previous studies manually labeled the dataset for training and testing purposes, the small amount of data is always considered to be a limitation of those studies. We hypothesize that conversational data that are similar to developer discussions might work as the training data we need. Hence, we sourced our data from \SO{} conversations in the form of questions, answers and comments which are already tagged by several developers and moderators \cite{Bazelli2013}. 
    
    Although we understand that the \SO{} conversations are not directly comparable with developer discussions, the words of those posts often contain architecture-relevant knowledge \cite{soliman2016}. Since this study is about classifying a discussion as design or non-design, conversational texts from \SO{} can provide those words that could be used to distinguish between design and non-design classes. We use this dataset only to train and validate our model but the actual testing of the model is conducted with the developer discussions dataset which we obtain from the study at \cite{Brunet2014}. This allowed us to obtain a dataset of 260,000 examples which is the largest dataset in design mining so far (the latest study from Viviani et al. \cite{Viviani2019} introduced a dataset of 2500 examples). 
    
 We explain the validity of this data in \S \ref{subsec:data_validation} and illustrate some of the improvements we notice in the results section in \S \ref{results-discussion}.
    \item \emph{How useful are software-specific word vectorizers?} \\
    \textbf{Approach---}Converting words of a text conversation to vectors as feature space representation is a common practice in Natural Language Processing. Previous studies have introduced various vectorization techniques. In response  to our previous study \cite{Mahadi_2020}, we demonstrate how word embedding as a vectorization choice can improve the performance of the classifier. However, word embedding needs a reference model. In the previous study, we used a general-purpose reference model that is trained on texts from Wikipedia. However, some of the software engineering context can get lost if we use general purpose reference model \cite{Efstathiou2018}. 
    
    For this reason, we decide to build our own reference model that is trained on software engineering related literature. We scrape the plain text from 300 books, conference and journal paper and develop a software-specific corpus to be used to train our software-specific word vectorizer.  We train our software-specific word embedding reference model based on the corpus and test it's performance and validity with respect to the general-purpose reference model to address research question 2. We explain the data collection method for this model elaborately in \S \ref{subsec:word_embedding_dataset} and the improvements in classification in \S \ref{subsec:software_specific_we_result}.
    \item \emph{How to provide domain context to a small sample of data?} \\
    \textbf{Approach---}To answer research question 3, we take every word from our training and testing data and inject similar words using techniques from \cite{Efstathiou2018} to augment the data we use for training and testing. Similar word models are unsupervised models trained on a corpus of text. They can output similar words of a word depending on the position and usage of that particular word with respect to the neighboring words. 
    
    We show an example of total-domain and cross-domain augmentation using similar word injection model. We use two word injection models: one from the train domain and the other from the test domain. We use augmentation for both the domain in order to transfer some of the context from each domain to another in the form of similar words. Finally, we demonstrate a new state of the art (SOTA) results in cross-domain design mining. We explain the design of our study for augmentation in \S \ref{subsec:swi} and discuss the state-of-the-art results in \S \ref{subsec:data_augmentation_result}.
\end{enumerate}

\noindent In this paper, we contribute the following:
\begin{itemize}
    \item We provide a labeled data set of two hundred and sixty thousand discussions in the form of train, test and validation data. This data set is fully processed with state-of-the-art and modern NLP standard and convention. We make this available in our replication packages at \replicationUrl{}.
    \item We present our software specific word vectorizer trained on hundreds of well processed and spell corrected literature on software engineering.
    \item We present, integrate and discuss two similar word injector models and show how to achieve total and cross domain context with them.
    \item We report on the performance of several machine learning models based on our approach.
\end{itemize}

This work extends our previous work \cite{Mahadi_2020}. The major extensions in this paper are as follows:
\begin{enumerate}
  \item We introduce a new dataset using data from \SO{} tagged with specific design keywords.
  \item We develop a software-specific corpus based on 300 software engineering texts.
  \item We build a software-specific word vector to test the value of context and augmentation in our approaches, challenges identified in the previous paper.
  \item We demonstrate new state of the art (SOTA) results in cross-project design mining.
\end{enumerate}

Our paper begins with a strict replication of the 2014 work of Brunet et al. \cite{Brunet2014} as a way of explaining the design mining research problem (\S \ref{sect:strict_replication}). We then extend the replication by examining improved techniques for dealing with the problem, including accounting for class imbalance, in \S \ref{sect:extending_the_replication}. We briefly overview our use of a deep learning language model, ULMFiT, and our attempt to do cross-project/cross-domain transfer learning in \S \ref{sect:conclusion_s}. 
We then go into detail about the challenges we faced with this transfer learning (\S \ref{dataset-design-discussion}) and then explain how we dealt with the issue of insufficient labeled data, and the need for software specific context, in \S \ref{sec:solns}. Our study design is explained in \S \ref{sec:design} and final results in \S \ref{results-discussion}. We finish the paper by characterizing some limitations and study design issues. 

%% file: sections/02-background-related.tex
\section{Background and Related Work}
\label{sect:related_work}

Our paper brings together two streams of previous research. First,  we highlight work on cross-project prediction and learning in software engineering.
Secondly, we discuss previous work in mining design discussions and summarize existing results as an informal meta-analysis.
We conclude by looking at the challenges of degrees of freedom in this type of research. 

\subsection{Cross-Project Classifiers in Software Engineering}
\label{subsec:cross_project_classifier}
A practically relevant classifier is one that can ingest a text snippet---design discussion---from a previously unseen software design artifact, and label it \textsf{Design/ Not-Design} with high accuracy. 
Since the classifier is almost certainly trained on a different set of data, the ability to make cross-dataset classifications is vital. 
Cross-dataset classification \cite{Zimmermann:2009} is the ability to train a model on one dataset and have it correctly classify other, different datasets. This is most important when we expect to see different data if the model is put into production. It might be less important in a corporate environment where the company has a large set of existing data (source code, for example) that can be used for training.

The challenge is that the underlying feature space and distribution of the new datasets differ from that of the original dataset, and therefore the classifier often performs poorly. 
For software data, the differences might be in the type of software being built, the size of the project, or how developers report bugs. Herbold \cite{herbold2017} conducted a mapping study of cross-project defect prediction which identified such efforts as strict (no use of other project data in training) or mixed, where it is permissible to mix different project data. We will examine both approaches in this paper, but in the domain of design mining, not defect prediction. Recent work by Bangash et al. \cite{bangash20} has reported on the importance of time-travel in defect prediction. Time-travel refers to the bias induced in training when using data from the future to predict the past. We do not think temporality is an important concern for the strict design mining problem we discuss here, but it might be relevant if one were to examine design evolution. In the case of design mining, we are not making predictions, and therefore the label results reflect an atemporal, holistic view of the project's total state of design. It is future work to capture how design is changing over time, where temporality might play a role.

To enable cross-domain learning without re-training the underlying models, the field of transfer learning applies machine learning techniques to improve the transfer between feature spaces \cite{Pan:2010}. Typically this means learning the two feature spaces and creating mapping functions to identify commonalities.
There have been several lines of research into transfer learning in software engineering. We summarize a few here. 
Zimmermann et al. \cite{Zimmermann:2009} conducted an extensive study of conclusion stability in defect predictors. 
Their study sought to understand how well a predictor trained with (for example) defect data from one brand of web browser might work on a distinct dataset from a competing web browser. 
Only 3.4\% of cross-project predictions achieved over 75\% accuracy, suggesting transfer of defect predictors was difficult. 

Following this work, a number of other papers have looked at conclusion stability and transfer learning within the fields of effort estimation and defect prediction. Herbold gives a good summary \cite{herbold2017}. 
Sharma et al \cite{Sharma:2019aa} have applied transfer learning to the problem of code smell detection. 
They used deep learning models and showed some success in transferring the classifier between C\# and Java. However, they focus on source code mining, and not natural language discussions. Code smells, defect prediction, or effort estimation are quite distinct from our work in design discussion, however, since they tend to deal with numeric data, as opposed to natural language.

Other approaches include the use of bellwethers \cite{Krishna:2016:TMA:2970276.2970339}, exemplar datasets that can be used as simple baseline dataset for generating quick predictions. The concept of bellwether for design is intriguing, since elements of software design, such as patterns and tactics, are generalizable to many different contexts.

Transfer learning in natural language processing tasks for software engineering is in its infancy. There is a lot of work in language models for software engineering tasks, but typically focused only on source code. Source code is highly regular and thus one would expect transferability to be less of a problem \cite{hindlenatural}. Early results from Robbes and Janes \cite{Robbes:2019} reported on using ULMFiT \cite{Howard:2018} for sentiment analysis with some success. We also use the transfer NLP potential of ULMFiT, which we discuss in \cite{Mahadi_2020}. Robbes and Janes emphasized the importance of pre-training the learner on (potentially small) task-specific datasets. We extensively investigate the usefulness of this approach with respect to design mining. Novielli et al. \cite{novielli20} characterize the ability of sentiment analysis tools to work without access to extensive sets of labeled data across projects, much as we do for design mining.

\subsection{Mining Design Discussions}
\label{sec:related_mining}
While repository mining of software artifacts has existed for two decades or more, mining repositories for \emph{design-related} information is relatively recent. In 2011 Hindle et al. 
proposed labeling non-functional requirements in order to track a project's relative focus on particular design-related software qualities, such as maintainability \cite{hindle11msr}. Hindle later extended that work \cite{HindleBZN15} by seeking to cross-reference commits with design documents at Microsoft. Brunet et al. \cite{Brunet2014} conducted an empirical study of design discussions, and is the target of our strict replication effort. They pioneered the classification approach to design mining: supervised learning by labeling a corpus of design discussions, then training a machine learning algorithm validated using cross-validation. 

Table \ref{table:related} (see Appendix) reviews the different approaches to the problem, and characterize them along the dimensions of how the study defined ``design'', how prevalent design discussions were, what projects were studied, and overall accuracy for the chosen approaches. We found 12 primary studies that look at design mining, based on a non-systematic literature search. We then conducted a rudimentary vote-counting meta-review \cite{Pickard1998} to derive some overall estimates for the feasibility of this approach (final row in the table).
 
\noindent\textbf{Defining Design Discussions}---The typical unit of analysis in these design mining studies is the ``discussion'', i.e., the interactive back-and-forth between project developers, stakeholders, and users. As Table \ref{table:related} shows, this varies based on the dataset being studied. A discussion can be code comments, commit comments, IRC or messaging application chats, Github pull request comments, and so on. The challenge is that the nature of the conversation changes based on the medium used; one might reasonably expect different discussions to be conducted over IRC vs a pull request.

\noindent\textbf{Frequency of Design Discussions}---Aranda and Venolia \cite{Aranda_2009} pointed out in 2009 that many software artifacts do not contain the entirety of important information for a given research question (in their case, bug reports). Design is, if anything, even less likely to appear in artifacts such as issue trackers, since it operates at a higher level of abstraction. Therefore we report on the average prevalence of design-related information in the studies we consider. On average 15\% of a corpus is design-related, but this is highly dependent on the artifact source. 
    
\noindent\textbf{Validation Approaches for Supervised Learning}---In Table \ref{table:related} column \textsf{Effectiveness} reports on how each study evaluated the performance of the machine learning choices made. These were mostly the typical machine learning measures: accuracy (number of true positives + true negatives divided by the total size of the labeled data), precision and recall (true positives found in all results, proportion of results that were true positives), and F1 measure (harmonic mean of precision and recall). 
Few studies use more robust analyses such as AUC (area under ROC curve, also known as balanced accuracy, defined as the rate of change). 
Since we are more interested in design discussions, which are the minority class of the dataset, AUC or balanced accuracy gives a better understanding of the result, because of the unbalanced nature of the dataset.

\noindent\textbf{Qualitative Analysis}
The qualitative approach to design mining is to conduct what amount to be targeted, qualitative assessments of projects. The datasets are notably smaller, in order to scale to the number of analysts, but the potential information is richer, since a trained eye is cast over the terms. The distinction with supervised labeling is that these studies are often opportunistic, as the analyst follows potentially interesting tangents (e.g., via issue hyperlinks). Ernst and Murphy \cite{Ernst:2012wf} used this case study approach to analyze how requirements and design were discussed in open-source projects. One follow-up to this work is that of Viviani, \cite{Viviani2018,Viviani2019}, papers which focus on rubrics for identifying design discussions. The advantage to the qualitative approach is that it can use more nuance in labeling design discussions at a more specific level; the tradeoff of course is such labeling is labour-intensive. 

\noindent\textbf{Summary}
A true meta-analysis \cite{Pickard1998,Kitchenham2019} of the related work is not feasible in the area of design mining. 
Conventional meta-analysis is applied on primary studies that conduct experiments in order to support inference to a population, which is not the study logic of the studies considered here. 
For example, there are no sampling frames or effect size calculations. 
One approach to assessing whether design mining studies have shown the ability to detect design is with vote-counting (\cite{Pickard1998}), i.e., count the studies with positive and negative effects past some threshold. 

As a form of vote-counting, the last row of Table \ref{table:related} averages the study results to derive estimates. 
On average, each study targets 29,298 discussions for training, focus mostly on open-source projects, and find design discussions in 14\% of the discussions studied. 

As for effectiveness of the machine learning approaches, here we need to define what an `effective' ML approach is. 
Since this is a two label problem a random guessing approach on a balanced dataset would achieve accuracy of 50\%. Balanced accuracy is defined as $$ BA =  \frac{TNR + TPR }{2}$$
where TNR = true negative rate (specificity), and TPR = true positive rate or sensitivity, aka recall. The balanced accuracy will always be 50\% for a random guesser in the 2 label case and is equivalent to the P/R AUC (area under the precision/recall curve) for that case.
From Table \ref{table:related} we see that the typical classifier in the literature can achieve close to 86\% performance, including Zanaty et al. \cite{Zanaty2018}, Viviani et al. \cite{Viviani2019}, and our previous paper \cite{Mahadi_2020}. Several other papers report higher scores but on imbalanced data. Furthermore, the best results reported in the table reflect algorithms largely tested on the same type of data (within dataset testing, such as IRC logs, pull requests, or issues). In this paper we tackle the challenge of cross-dataset accuracy.


\subsection{The Role of Researcher Degrees of Freedom}
A related issue to conclusion stability is the concept of researcher degrees of freedom (RDOF). RDOF \cite{Gelman:2013aa,Gelman:2012aa} refers to the multiple, equally probable analysis paths present in any research study, any of which might lead to a \emph{significant} result.  
Failure to account for researcher degrees of freedom directly impacts conclusion stability and overall practical relevance of the work, as shown in papers such as Di Nucci et al. \cite{DiNucci2018} and Hill et al. \cite{Hill2012}.
For example, for many decisions in mining studies like this one, there are competing views on when and how they should be used, multiple possible pre-processing choices, and several ways to interpret results. 
Indeed, the approach we outlined here in Figure \ref{fig:study_design} is over-simplified, given the actual number of choices we encountered. 
Furthermore, the existence of some choices may not be apparent to someone not deeply skilled in these types of studies. 

A related concept is the notion of \emph{conclusion stability} from Menzies and Sheppherd \cite{Menzies2012}. 
Conclusion stability is the notion that an effect X that is detected in one situation (or dataset) will also appear in other situations. 
Conclusion stability suggests that the theory that predicts an effect X holds (transfers) to other datasets. In design mining, then, conclusion stability is closely tied to the ability to transfer models to different datasets.

That led to work on the problem with bias and variance in sampling for software engineering studies \cite{Kocaguneli2013}, where Kocaguneli and Menzies concluded that perhaps this tradeoff is not as important in software studies\footnote{from \cite{Menzies2012}, where $S$ is a study accuracy statistic, and $\hat{S}$ is the population (true) statistic: ``bias is $S - E(\hat{S})$ where $E$ is the expected value and the variance is $E((S - \hat{S})^2)$"}. 

One possible approach is to use toolkits with intelligently tuned parameter choices. 
Hyper-parameter tuning is one such example of applying machine learning to the problem of machine learning, and research is promising \cite{xia18}. 
Clearly one particular analysis path will not apply broadly to \emph{all} software projects.
What we should aim for, however, is to outline the edges of where and more importantly, why these differences exist.

%% file: saner_sections/03-replication.tex
\section{Design Mining Replication}

\subsection{Strict Replication}
\label{sect:strict_replication}
We begin by showing how design mining works, by replicating the existing design mining studies and exploring the best combination of features for state of the art results. 
We conduct a strict replication (after Gómez et al. \cite{Gmez2014}), a replication with little to no variance from the original study, apart from a change in the experimenters. 
However, given this is a computational data study, researcher bias is less of a concern than lab or field studies (cf. \cite{Storey:2019aa}).
The purpose of these strict replications is to explain the current approaches and examine if recent improvements in NLP might improve the state of the art.

\begin{figure}[hbt]
\centering
  \includegraphics[width=0.45\textwidth]{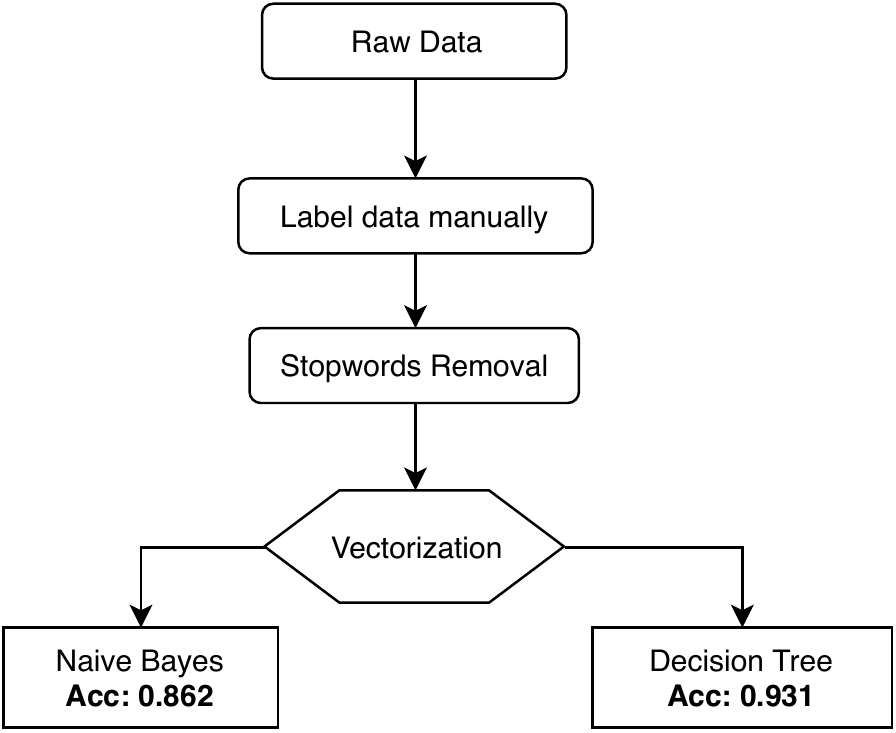}
  \caption{Protocol map of Brunet \cite{Brunet2014} study}
  \label{fig:brunet-protocol}
\end{figure}

To explain the differences in studies, we use protocol maps, a graphical framework for explaining an analysis protocol. This graphical representation is intended to provide a visual device for comprehending the scope of analysis choices in a given study. 
Fig. \ref{fig:brunet-protocol} shows a protocol map for the strict replication. 
The enumerated list that follows matches the numbers in the protocol diagram. 
\begin{enumerate}
  \item Brunet's study \cite{Brunet2014} selected data from 77 Github projects using their discussions found in pull requests and issues.
  \item Brunet and his colleagues labeled 1000 of those discussions using a coding guide.
  \item Stopwords were removed. They used NLTK stopwords dictionary and self defined stopsets.
  \item The data were vectorized, using a combined bigram word feature and using the NLTK BigramCollectionFinder to take top 200 ngrams.
  \item Finally, Brunet applied two machine learning approaches, Naive Bayes and Decision Trees. 10-fold cross validation produced the results shown in Fig. \ref{fig:brunet-protocol}: mean accuracy of \textbf{0.862} for NaiveBayes, and \textbf{0.931} for Decision Trees, which is also slower.
\end{enumerate}

We followed this protocol strictly. We downloaded the data that Brunet has made available; applied his list of stop words; and then used Decision Trees and NaiveBayes to obtain the same accuracy scores as his paper. The only difference is the use of \textsf{scikit-learn} for the classifiers, instead of NLTK. Doing this allowed us to match the results that the original paper \cite{Brunet2014} obtained.

We did notice one potential omission. 1000 sentences are manually classified in Brunet's dataset \cite{Brunet2014}. However, only 224 of them are design, which indicates serious imbalance in the data. As a result, the accuracy measure, which assumes a balanced set of classes, likely overstates the true validity of this approach.

\begin{figure*}[hbt]
\centering
  \includegraphics[width=0.9\textwidth]{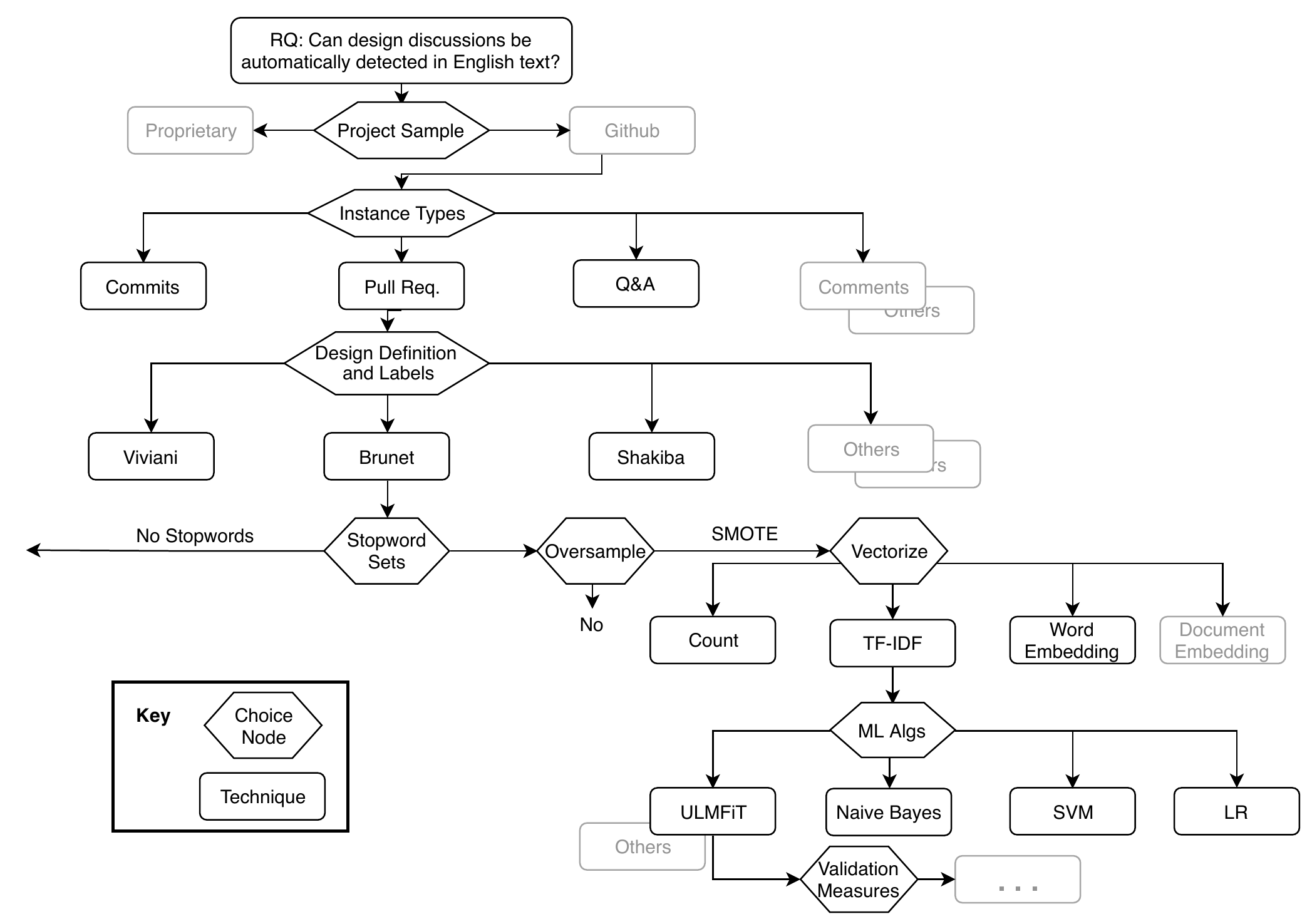}
  \caption{Protocol map of possible research paths for design mining studies.}
  \label{fig:extensions}
\end{figure*}

\subsection{Extending the Replication}
\label{sect:extending_the_replication}

A strict replication is useful to confirm results, which we did, but does not offer much in the way of new insights into the underlying research questions. In this case, we want to understand how to best extract these design discussions from \emph{any} corpora. This should help understand what features are important for our goal of improving conclusion stability.

Shepperd \cite{Shepperd2018} shows that focusing (only) on replication ignores the real goal, namely, to increase confidence in the result. Shepperd's paper focused on the case of null-hypothesis testing, e.g., comparison of means. In the design mining problem, our confidence is based on the validation measures, and we say (as do Brunet and the papers we discussed in \S \ref{sec:related_mining}) that we have more confidence in the result of a classifier study if the accuracy (or similar measures of classifier performance) is higher. 

However, this is a narrow definition of confidence; ultimately we have a more stable set of conclusions (i.e. that design discussions can be extracted with supervised learning) if we can repeat this study with entirely different datasets. We first discuss how to improve the protocol for replication, and then, in Section \ref{sect:conclusion_s}, discuss how this protocol might be applied to other, different datasets.

We extend the previous replication in several directions. Fig. \ref{fig:extensions} shows the summary of the extensions, with many branches of the tree omitted for space reasons. The complete extension can be found at \extensionFigureUrl.
One immediate observation is that it is unsurprising conclusion stability is challenging to achieve, given the large researcher degrees of freedom, i.e., number of analysis choices a researcher could pursue.
We found several steps where Brunet's original approach could be improved. These improvements could also apply to other studies shown in Table \ref{table:related}. 

We switched to use balanced accuracy, or area under the receiver operating characteristic curve (AUC-ROC or AUC), since it is a better predictor of performance in imbalanced datasets\footnote{defined in the two-label case as the True Positive Rate + the False Positive Rate, divided by two}.


\begin{figure}[hbt]
\centering
  \includegraphics[width=0.35\textwidth]{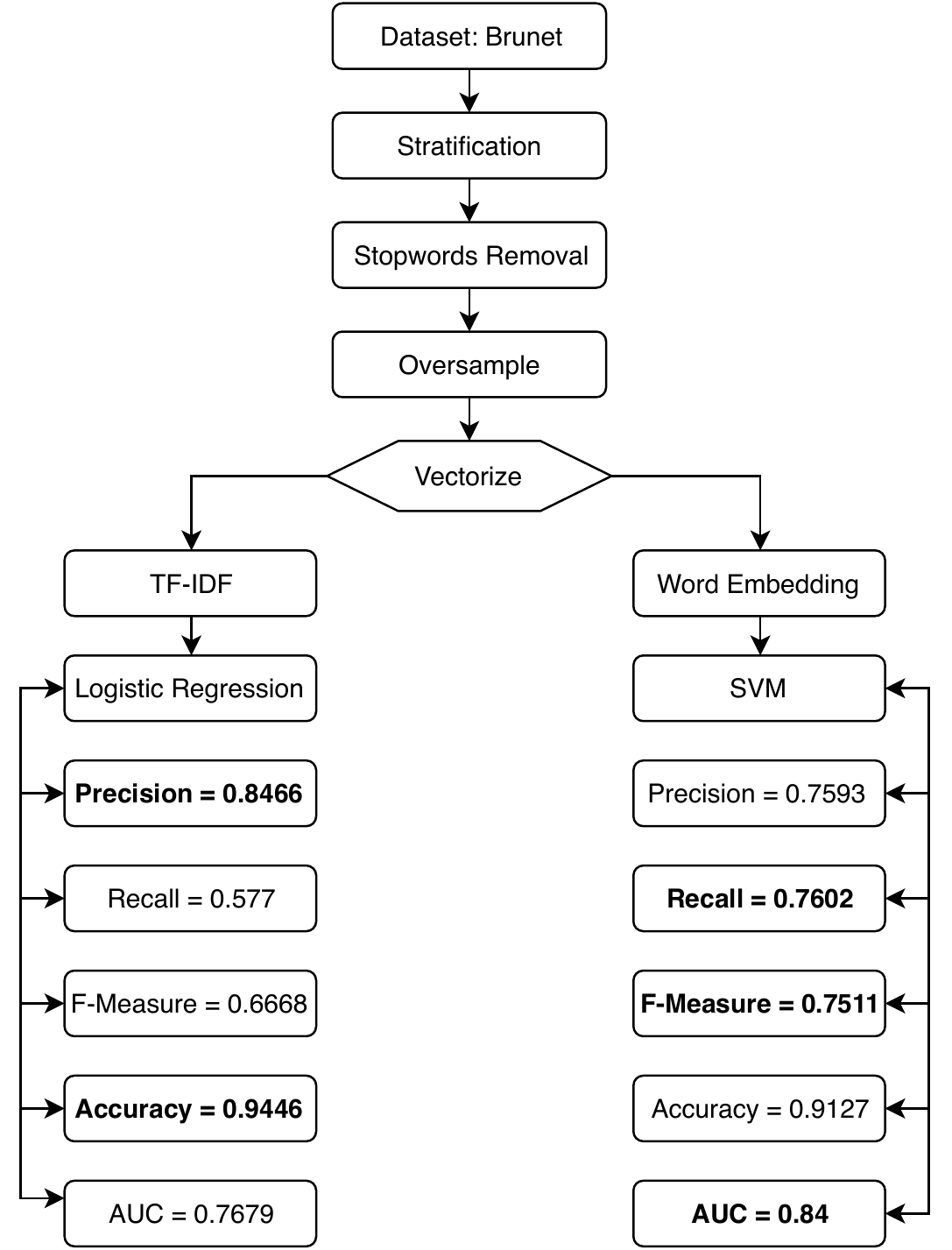}
  \caption{Preferred Design Mining Method \textsf{NewBest}. Numbers are mean of 10-fold cross validation.}
  \label{fig:preferred_method}
\end{figure}

\subsubsection{Vectorization Choices}
\label{sec:wordembed}
Vectorization refers to the way in which the natural language words in a design discussion are represented as numerical vectors, which is necessary for classification algorithms. We present four choices: one, a simple count; two, term-frequency/inverse document frequency (TF-IDF), three, word embeddings, and four, document embeddings. The first two are relatively common so we focus on the last two.
%
%

Word embeddings are vector space representations of word similarity. Our intuition is this model should capture design discussions better than other vectorization approaches. A word embedding is first trained on a corpus. In this study, we consider two vectorization approaches, and one similarity embedding. ``Wiki'' is a Fasttext embedding produced from training on the Wikipedia database plus news articles \cite{mikolov2018advances}, and GloVe, trained on web crawling \cite{pennington2014glove}. The final embedding is trained on the \SO{} dataset, courtesy of Efstathiou et al. \cite{Efstathiou2018}. While Wikipedia considers more words in English, the \SO{} dataset should be more representative of the software domain. The embedding is then used to either a) train a classifier like Logistic Regression by passing new discussions to the embedding, and receiving a vector of its spatial representation in return; b) expanding the scope of small discussions by adding related words to the sentence (\SO{}).

As Fig. \ref{fig:extensions} shows, there are several ways in which vectorization applies. We also wanted to see if we could expand the size of the training set by using the in-built capability of a word embedding to identify similar words (i.e., words that are close in vector space). The intuition is that since the discussions from the Brunet dataset are typically quite short, we could add similar words  to extend the vocabulary. However, the vocabulary expansion approach did not make any difference to our accuracy results, likely because domain-specific terms like ``library" were overwhelmed with standard English terms like ``thus|hence''.

\subsubsection{Document Vectors and the \SO{} Design Corpus}
\label{sec:doc2vec}
Extending word vectors, we can also capture the spatial representation of entire discussions. 
Document vectors, introduced in \cite{doc2vec14}, are extensions to word embeddings that add an extra dimension to the vector space to capture the document of origin (in this case, a design discussion). 
The approach has been shown to improve the ability to capture discussion-wide meaning, where a word embedding approach alone would be focused on a smaller window of words. 
We used the Gensim Doc2Vec class to train a document embedding on 26,969 \SO{} questions and answers, that were tagged with the label `design' (which we extracted from the SOTorrent dataset of Baltes et al. \cite{BaltesDT008}), combined with 25,000 random questions \emph{not} tagged design.\footnote{This dataset can be found as part of our replication package}

We processed the data to remove stopwords, HTML and \verb| <code>| tags (including the code snippets found within). 
We also removed graphical or web design discussions, where they had tags that co-occurred with the `design' tag, such as \texttt{CSS,HTML}. 
We then trained a document vector based on the 51,969 documents in the corpus, and used logistic regression to classify the documents as either design or not. 
Results are good on the internal, within-sample classification, even for the test set: accuracy of 0.934 in training, and 0.932 for test (held-out) data. This dataset is balanced so accuracy is a reasonable validation metric.


\subsubsection{Other Extensions}

We used imbalance correction in order to account for the fact design discussion make up only 15\% (average) labels. We took two approaches.
One, we stratified folds to keep the ratio of positive and negative data equal. 
After stratifying, we have again run the experiment described in \cite{Brunet2014} and examined that the accuracy dropped significantly from reported 94\% to around 87.6\% where our experiment achieved an accuracy of around 94\%. 
We use SMOTE \cite{Chawla:2002aa} to correct for imbalanced classes in train data. Recall from Table \ref{table:related} that design discussion prevalence is at best 14\%. This means that training examples are heavily weighted to non-design instances. As in \cite{Alkadhi2017}, we correct for this by increasing the ratio of training instances to balance the design and non-design instances. We have oversampled the minority class (i.e., `design'). 

We also hypothesized that the software-specific nature of design discussions might mean using non-software training data would not yield good results. Specifically, when it comes to stopword removal, we used our own domain-specific stopword set along with the predefined English stopwords (of scikit-learn). We also searched for other words that may not mean anything significant, such as `lgtm' (`looks good to me') or `pinging', which is a way to tag someone to a discussion. These stopwords may vary depending on the project culture and interaction style, so we removed them. 

\subsubsection{Best Performing Protocol}
After applying these extensions, Fig. \ref{fig:preferred_method} shows the final approach. Ultimately, for our best set of choices we were able to obtain an AUC measure of 0.84, comparable to the unbalanced accuracy Brunet reported of 0.931. The Matthews correlation coefficient (MCC), which measures correlation between true labels and actual labels, was 0.63.

Logistic Regression with TF-IDF vectorization gives the best results in terms of Precision and Accuracy. 
On the other hand, Word Embedding with Support Vector Machine provides best results in terms of Recall, F-Measure and Balanced Accuracy or AUC. 
Since we are interested in the `design' class which is the minority class of the dataset, highest Recall value should be more acceptable than Precision. 
As a result we created a \textsf{NewBest} classifier based on the combination  of `Word Embedding' and `Support Vector Machine' (right hand of Fig. \ref{fig:preferred_method}).

%% file: sections/deep_approach.tex
\begin{table*}[h]
 \centering
 \caption{Datasets used for within and cross-dataset classification. All datasets are English-language}
 \label{tbl:newdatasets}
 \begin{tabular}{p{0.9cm}p{0.9cm}p{1.3cm}p{1cm}p{0.8cm}p{2cm}p{1cm}p{1cm}} 
 \toprule
 \textbf{Citation} & \textbf{Dataset} & \textbf{Type}  & \textbf{Total instances} & \textbf{Design instances} & \textbf{Projects} & \textbf{Mean Discussion Length (words) } & \textbf{Voca-bulary Size (words)} \\
 \midrule
 \cite{Brunet2014} & Brunet 2014  &  Pull requests & 1,000 & 224 & BitCoin, Akka, OpenFramework, Mono, Finagle & 16.97 & 3,215 \\
\cite{Shakiba2016} & Shakiba 2016  & Commit messages & 2,000 & 279 & Random Github and SourceForge  & 7.43 & 4,695 \\
 \cite{Viviani2018} & Viviani 2018   &   Pull requests & 5,062 & 2,372 & Node, Rust, Rails & 36.13 &  24,141 \\
\cite{Maldonado2017} & SATD  & Code comments & 62,276 & 2,703 & 10 Java incl Ant, jEdit, ArgoUML & 59.13 & 49,945 \\ 
\cite{Mahadi_2020} & \SO{} & \SO{} questions & 51,989 & 26,989 & n/a & 114.79 & 252,565 \\
 \bottomrule
 \end{tabular}
 \end{table*}

 \section{Testing Cross-Project Conclusion Stability}
\label{sect:conclusion_s}
In this section we build on the replication results and enhancements of our strict replication. We have a highly accurate classifier, NewBest, that does well \emph{within-dataset}. We now explore its validity when applied to other datasets, i.e., whether it has conclusion stability. 

In \cite{Menzies2012}, Menzies and Shepperd discuss how to ensure conclusion stability. They point out that predictor performance can change dramatically depending on dataset (as it did in Zimmermann et al. \cite{Zimmermann:2009}). Menzies and Shepperd specifically analyze prediction studies, but we believe this can be generalized to classification as well. Their recommendations are to a) gather more datasets and b) use train/test sampling (that is, test the tool on different data entirely). 

In this section we evaluate a classifier trained on one dataset to a different dataset, but consisting of the same types of discussions. Before beginning to apply learners to different datasets, it makes sense to ask if this transfer is reasonable. For example, in Zimmermann et al. \cite{Zimmermann:2009} the specific characteristics of each project were presented in order to explain the intuition behind transfer. 
E.g., should a discussion of design in \SO{} be transferable, that is, considered largely similar to, one from Github pull requests? While the artifacts are different, and used in different circumstances, we believe they both contain design information (from preceding studies), and are natural language discussions.
 
In Table \ref{tbl:newdatasets} we illustrate each of the datasets considered in this paper. In Table \ref{tbl:examples} we show some sample design discussions from each.
Since performance of transfer learning is largely based on similarity between projects (i.e., feature spaces), we would expect to see better AUC results for cross-project prediction if data sources are the same (e.g. pull requests), projects are the same, and/or the platforms are the same (e.g. Github). 

\subsection{Research Method}
We test the ability to transfer classifiers to new types of discussions and datasets.
We applied the best protocol result from above. That is, the NewBest classifier, using \textsf{stopwords+oversampling+TF-IDF+Logistic Regression}. 
We train this classifier on the Brunet \cite{Brunet2014} data, and the other 4 datasets described in Table \ref{tbl:newdatasets}.

We then apply the trained model, as well as the ULMFiT model described below, to each dataset in turn (thus, 5 comparisons, including within-project labeling for a baseline).

 \begin{table}[htb]
 \centering
 \caption{Sample (raw) design discussions, pre data cleaning.}
 \label{tbl:examples}
 \begin{tabular}{cp{6.2cm}} 
 \toprule
 \textbf{Dataset} & \textbf{Sample Snippet}  \\
 \midrule
\SO{} & \emph{What software do you use when designing classes and their relationship, or just pen and paper?} \\
Brunet 2014 & \emph{Looks great Patrik Since this is general purpose does it belong in util Or does that introduce an unwanted dependency on dispatch} \\
SATD & \emph{// TODO: allow user to request the system or no parent}\\
Viviani 2018 & \emph{Switching the default will make all of those tutorials and chunks of code fail with routing errors, and ``the RFC says X" doesn't seem like anywhere near a good enough reason to do that.} \\
Shakiba 2016 & \emph{Move saveCallback and loadCallback to RequestProcessor class } \\
 \bottomrule
 \end{tabular}
 \end{table}

\subsection{Transfer Learning with ULMFiT}
\label{sec:ulmfit}
Early results reported by Robbes and Janes \cite{Robbes:2019} suggested recent work on ULMFiT (Universal Language Model Fine-Tuning, \cite{Howard:2018}) might work well for transfer learning in NLP for the software domain. They applied it to the task of sentiment analysis. 

 ULMFiT uses a three layer bi-LSTM (long short-term memory) architecture. It supports transfer learning for NLP tasks without having to train a new model from the beginning. ULMFiT uses novel NLP techniques like discriminative fine tuning, gradual unfreezing and slanted triangular learning rates which makes it state-of-the-art \cite{Howard:2018}. 
 ULMFiT comes pre-trained using data on a Wikipedia dataset. Wikipedia lacks software specificity.
 Our aim using ULMFiT was to observe the behavior of the pre-trained ULMFiT model (AWD-LSTM) when we train its last neural network layer with the \SO{} design discussion data. The \SO{} data fine-tunes the contextual layers of the model.
 We combined the \SO{} data along with the four design-specific datasets for training. 

For our ULMFiT deep learning (DL) approach we used the DL practice \cite{James2013} of a 60\%-20\%-20\% train-validate-test ratio, where the test set is held back from training and validation. We ran this 3 times and noticed only trivial changes (less than 0.5\%) in the results for any of the runs. We report the mean.

We had two reasons for this choice. First, the anticipated training costs and data size are much larger in deep network models. Second, we wanted to test the claimed capability of ULMFiT \cite{Howard:2018} to performs well on small sample sizes (60\% of our dataset). Unsurprisingly, adding more data points increases performance, at the cost of overfitting (detailed results can be found in our replication package).

Fine-tuning ULMFiT incorporates internal error assessment using a validation and train loss technique to determine the optimal trained model by constantly observing the difference between train loss and validation loss. 
This informs model selection, which is then tested against the held-back test set.

Training ULMFiT involves first, model pre-training using the AWD-LSTM state of the art language modelling technique; second, fine tuning the learning rate of the language model to get the optimal value of learning rate. This can be done per layer of the neural network.

In the third stage, we train the language classification model on top of a pre-trained language learner model. The training data remains \SO{}, but now in the supervised, design-labeled context. Finally, we again fine-tune our trained text classifier learner to find an optimal learning rate to gain a good balance between overfitting and underfitting the model. For this experiment, our optimal learning rate was 0.01. 

\begin{figure}[hbt]
\centering
  \includegraphics[width=0.75\textwidth]{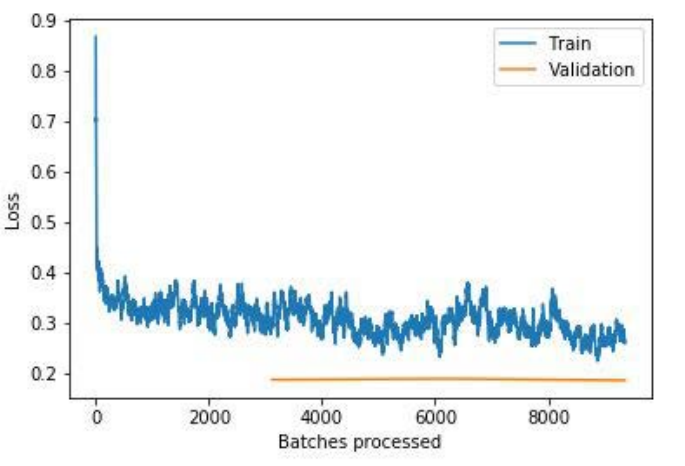}
  \caption{Loss plot after 3 epochs}
  \label{fig:Lossplot}
\end{figure}

After performing the above steps by combining the \SO{} dataset---questions tagged design/non-design---and the Brunet2014 dataset, AUC was approximately 93\% during the training phase (i.e., within sample performance).
To ensure the model is not overfitting or underfitting, we plot the recorded train and validation losses. 
We make sure that the train and validation losses are close to each other. 
Fig. \ref{fig:Lossplot} is the result after 3 epochs of model training with the learning rate of 0.01, which is the optimal learning rate for our model. 
We see that the training loss is close to the validation loss (0.18 vs 0.22). 
This suggests the trained model is not over fitting. 


 \begin{figure}[hbt]
  \begin{subfigure}{.7\textwidth}
  \centering
  \includegraphics[angle=-90,width=\linewidth]{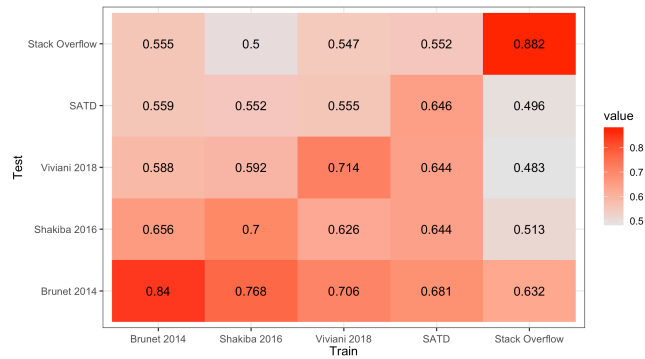}
  \caption{NewBest results.}
 \label{fig:sfig1}
\end{subfigure}

  \begin{subfigure}{.7\textwidth}
  \centering
  \includegraphics[angle=-90,width=\linewidth]{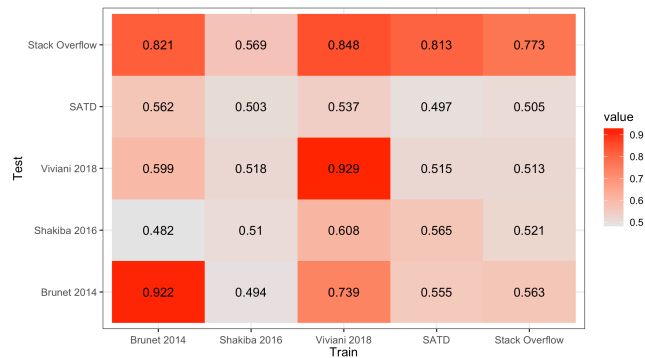}
  \caption{ULMFiT results. }
  \label{fig:sfig2}
\end{subfigure}

\caption{Cross-dataset design mining. Numbers: AUC. Read these plots as ``the model trained on the Dataset  on the X axis has AUC \emph{value} Tested On Dataset on the Y Axis". Higher intensity = better score.}
\label{fig:cross-results}
\end{figure}
%

Our current approach is trained using a LSTM Neural Network. This indicates there is also scope for fine-tuning several layers of the neural network in order to gain better performance on predictions. 

\subsection{Results}
Results are summarized in the heat maps shown in Fig.  \ref{fig:cross-results}. More intense color is better. Fig. \ref{fig:sfig1} shows the results for the NewBest protocol (SVM with word embeddings). Fig \ref{fig:sfig2} shows the equivalent for the ULMFiT approach. Our replication package includes complete results including confidence intervals and tests of significance.

The main challenge for conclusion stability with design mining datasets is that it is hard to normalize natural language text. 
This means while two datasets might reasonably be said to deal with design, one might have chat-like colloquial sentences, while the other has terse, template-driven comments.  We illustrate this difference with the example discussions shown in Table \ref{tbl:examples}.
In comparing to other datasets the comparison should still be over reasonably similar `apples'. As Table \ref{tbl:newdatasets} shows, there is some variance in all five datasets, with the type of discussion artifact, source projects, and linguistic characteristics differing. However, despite these differences Table \ref{tbl:examples} suggests there should be broad similarities: e.g., concepts such as \texttt{Class} or \texttt{User}, or ideas like moving functionality to different locations. Intuitively, we suggest the notion of transfer ought to work to some extent on these datasets: they are not completely different.

For the NewBest approach, the diagonal starting bottom-left captures the \emph{within-dataset} performance, which as expected, is better than the cross-dataset AUC scores. Secondly, all models performed best on the Brunet test dataset (bottom row). This is because in building the NewBest classifier, we evaluated our protocol choices against the Brunet dataset. This shows how tightly coupled protocol choices and conclusion stability are.

It also seems to be the case that results are poorer for datasets that are more removed from each other: using pull requests (Viviani and Brunet) does little better than random for \SO{} and code comments (SATD).

For the ULMFiT results in Fig. \ref{fig:sfig2}, we can see the benefit of training with the \SO{} dataset plus the fine-tuning (i.e., each tic on the x-axis reflects the fine-tuning in addition to \SO{} training). This is shown by the good results in the top row, which are essentially within-dataset results. Other results, however, are poor, and particularly when compared to the NewBest results (which is also much quicker to train). A Mann-Whitney test of significance (i.e., that the accuracy measures of ULMFiT are different than those from NewBest) shows no differences between the two learners (W = 356.5, p-value = 0.399). We therefore do not see much benefit from the way we have applied ULMFiT for cross-dataset classification of design. 

We ran 10-fold cross-validation on ULMFiT as well, and report those results in the replication package. For the transfer learning research objective, ten fold validation does not make a significant difference in AUC. Transfer performance on Brunet2014 data only changes from 0.502 to 0.588, even with the additional data. Thus, our conclusion that ULMFiT is not a significant improvement on the transfer learning problem remains the same.

Although the design discussions were in natural language, there were many words that were unique to the software/open source domain. We observed that the more vocabulary we feed to our ULMFiT model, the more it knows, the more it gets tuned on the language modelling, the better it performed. Thus we see great benefit in increasing the software-specific, dataset agnostic data we train ULMFiT.

%% file: sections/03-dataset-design-discussion.tex
\section{Challenges with Cross-Dataset Classification in Design Mining}
\label{dataset-design-discussion}

In Section \ref{sect:extending_the_replication} we introduced a design mining classifier that modified previous efforts by adding some modifications to the algorithm, and improving the ways of validating the classifier by implementing appropriate validation criteria to account for the imbalanced nature of the data. We also looked at using deep language models based on ULMFit to explore if fine-tuned deep models work well. Our underlying goals were aimed at improving the conclusion stability of a given model, namely, the model's performance on out of sample data, which might be from different projects (cross project) or across domain (different domains, such as Github issues, pull requests, source code comments, or \SO{} discussions).

For cross-dataset and cross-domain classification, our efforts performed poorly. Our classifier trained on one set of data does not perform particularly well in classifying discussions of other domains, such as \SO{} questions. Even though the preliminary results of Robbes and Janes \cite{Robbes:2019} suggested fine-tuning ULMFiT would do well out of sample, we did not see any benefit to this approach. 

After our evaluation with this deep learning model, we did not observe any significant improvement in the cross-dataset/domain learning problem. We found many unique words closely related to software domain. As a result, even though we were considering our problem to be related to typical NLP (Natural Language Processing), models trained on natural literature, such as Wikipedia, often failed to understand the context of software design. Thus our results are not surprising, since lack of domain specificity is a well-known limitation of such language models \cite{Efstathiou2018}.

We observed the following challenges:
\begin{enumerate}
\item As with all machine learning efforts, there is a linear increment on the learning with the amount of vocabulary. Our experiment suffered from a relative lack of data. Thus, it is necessary to find \emph{scalable ways of obtaining labeled design data}. 
\item Machine learning models trained on one domain get biased to that input format. For example, the way in which one writes an issue comment is different than how one writes a code comment (above and beyond the syntactical issues).  Thus we also concluded that we need to \emph{provide the classifier with the context of both domains} (issues, code comments, etc.) in order to transfer the learning from one domain to another.     
\item NLP models perform well with general discussions, however, they \emph{fail to understand the context of software specific vocabularies} \cite{karampatsis20}. For example, the word ``class'' has a completely different context in software engineering compared to general purpose literature.
\end{enumerate}

We therefore focus the remainder of this paper on dealing with these challenges in the context of design mining. 

\section{Solutions to the Challenges}
\label{sec:solns}
\subsection{Getting More Labeled Data}

Most design mining studies e.g., \cite{Brunet2014,Shakiba2016,Viviani2018b}, build a training dataset using human labellers, where two or more humans first label the data and then discuss among themselves to come to an agreement about the label (i.e., design/not-design). While we think this is a very effective method of building the dataset, the limitation on the amount of data a human can label is a problem that was repeatedly mentioned in the studies from before. This was also one of the conclusions we made in our previous study \cite{Mahadi_2020}. 

In this section we talk about our proposed idea about acquiring, processing and validating data that exist in different developer communication platforms. We are mainly building and leveraging two kinds of dataset in our study: one for our word embedding model (i.e., to vectorize text) and the other for the classification problem.

\subsubsection{Datasets}
In the following experiments, we report results using new datasets and approaches as they perform on the Brunet2014 data, as one possible cross-domain classification problem. We use data from the study of Brunet et at. \cite{Brunet2014} as our \textbf{cross-domain test data sample} and refer to this as Brunet2014 data for the rest of this paper. Brunet's data was initially extracted from Github, which fulfills our criteria of being a cross domain dataset. Note that we also have \textbf{within-dataset} results which we expect to be higher than the cross-domain results. We report this at the end of Section 8.2.

\paragraph{Dataset for Word Embedding}
\label{subsec:word_embedding_dataset}
We create a software-specific word embedding using software engineering literature.
The lack of semantics can harm the performance of a word embedding model \cite{Wang2016}. Word embeddings are distributed word representations based on neural networks. While traditional one-hot algorithm represents a word with a large vector, word embeddings embed every word into a low dimensional continuous space taking the semantic and syntactic information in account \cite{li2015}. 

Hence, our word embedding model will benefit most from structured sentences with grammatical correctness providing semantic relationship information between key terms \cite{Jian2008}. The software engineering literature represents more structured sentences in terms of semantics than comments or discussion threads. In our effort to get structured textual data, we scrape the plain text from 300 books, conferences and journal papers. We remove the individual title, figures, names, tables and all the meta-data to preserve the copyright policies of the literature. Our specific processing approach is discussed later. After all the processing, our dataset yields 1,575,439 total words with 20,607 unique words. We make this dataset and the following datasets available in our replication package at \replicationUrl.

\paragraph{Dataset for Classifier}
\label{subsec:classifier_dataset}
The word embedding can be used to vectorize an input into a format usable by machine learning algorithms. But what types of training data should that vectorization use?

We scrape text from \SO{} questions, answers and comments to use them as our training, testing and validation data for our classifier. For labeling the data, we use the user-assigned tags of the questions and label these tags as \textsf{design} or \textsf{general} depending on the tags of the question. Bazelli et al. \cite{Bazelli2013} showed that a tag acts as a label that can be used to describe the contents of the questions. Their study found that the tags can be at times ``misleading'', because they can be assigned both by the author of the questions as well as other users. However, they mostly represent actual information about the content of the question because of the moderation by designated moderators and removal of unused of tags after a certain period. 

If we find one of the following: ``design-patterns'', ``software-design'', ``class-design'', ``design-principles'', ``system-design'', ``code-design'', ``api-design'', ``language-design'', ``dependency-injection'', ``architecture'' tags, we label the entire discussion as ``design'' while labeling ``general'' otherwise. 

Table \ref{table:label_example} shows some output examples based on the code-book we defined, after text processing described in subsection \ref{subsec:data_processing}. 
The data distribution at Table \ref{table:data_distribution} illustrates that we have achieved a large number of data for our training, validation and testing phase with equal amount of instances for each class, preventing the problem of class imbalance for this type of data reported in \cite{Shakiba2016}. This even distribution of the dataset also removes some of the steps we took in \cite{Mahadi_2020} in our attempt to remove the unbalanced nature of the data such as stratification \cite{Sechidis} and oversampling \cite{Chawla:2002aa}.

\begin{table}[!ht]
    \centering
    \caption{Example of labeling \SO{} discussions based on tags}
    \begin{tabular}{|p{0.5\textwidth}|p{0.2\textwidth}|p{0.2\textwidth}|}
        \hline
        Text & Tags & Label \\
        \hline
        headless device local network trying headless raspberry connect local network want automated though mobile flutter given mobile network raspberry connected flutter will connected firebase & python, flutter, networking, dart, raspberry-pi & general \\
        \hline
        difference interface design pattern hard time knowing when something interface design pattern example observer & design-patterns, model-view-controller, interface, observer-pattern & design \\
        \hline
        soap request explain doing something platform called section called repeaters send soap request specific address will honest idea soap something question receive soap data want receive soap request know save works structure guys give information send soap imagine tried investigate truth know soap works using code seem work hope help thanks & php, xml, web-services, soap & general \\
        \hline
        behind naming visitor pattern book design pattern says visitor pattern visitor lets define changing classes elements read pattern book failing understand intuition behind naming pattern visitor called visitor & design-patterns, visitor & design \\
        \hline
    \end{tabular}
    \label{table:label_example}
\end{table}

\begin{table}[!ht]
    \centering
    \caption{Data Distribution for the Classifier}
    \begin{tabular}{cccc}
        \toprule
        Type & Total & \# Design data & \# General data \\ \midrule
        Train & 200,000 & 100,000 & 100,000 \\
        Validation & 30,000 & 15,000 & 15,000 \\
        Test & 30,000 & 15,000 & 15,000 \\
        \bottomrule
            \end{tabular}
    \label{table:data_distribution}
\end{table}

\subsubsection{Data Processing}
\label{subsec:data_processing}
Hemalatha et al.~\cite{Hemalatha2012} showed that data processing helps to remove noisy and inconsistent data resulting in improved performance, inspiring us to take a pipeline of different text processing techniques to process the data. Raw text from comments and discussions from the web pages often contains unnecessary tags, punctuation, white spaces, new lines and numeric elements. At the beginning of the processing, we look for these and strip them from the text. We implement our spell correction algorithm as the next step of the process. We take the Britain English dictionary with the affix as our primary dictionary and add Australian, Canadian, American and South African English vocabulary into it to make it more versatile. Then we take every misspelled word and compare it with the dictionary to find out the most relevant word. We take up to five relevant words and then analyzing them based on the context, number of similar letters present and the structure. Then we replace the misspelled word with the correct word or a list of possible correctly spelled words. 

Ghag et al. \cite{Ghag2015} show that stopword removal can significantly improve the performance of the traditional models, however it does not do much good to the more sophisticated deep learning based models. Since we are trying to make our dataset general and independent of a particular classifier, we take the stopword removal as our next step of processing the text. 
This step removes all the insignificant and unwanted words from the data such as ``the'', ``and'' etc. which in turn, can hamper the performance by misleading the classifier. We also make sure to keep our word length between 3 and 25 because any word less than 3 letters or more than 25 letters is not relevant to our study. 
At the final step of our processing, we implement a normalization approach to normalize the words to their root/base form. For example, `running' has a base word of run-. We choose lemmatization over stemming because Balakrishnan et al., \cite{Balakrishnan2014} shows some versatile experiments where lemmatization outperforms stemming in terms of precision. A base vocabulary is used to perform lemmatization which eliminates the inflectional endings and reduces the word to its base form. We use a lexical database called WordNet \cite{Miller1998} as our base vocabulary. Combining all the steps in a pipeline allow us to achieve a state-of-the-art text processor for our study.

\subsubsection{Data Validation}
\label{subsec:data_validation}
Section \ref{subsec:classifier_dataset} discussed how we acquire the data and label them with the help of user-created tags. However, tags can be misleading too. Often, tags are created by the author and the concept of design being very subjective, labeling data with the help of tags can create doubt on the validity of the data. In this section, we examine the validity of the data with the help of top words and top phrase analysis.

The presence of a particular word can contribute a lot in classifying the whole text. For example, if certain text contains keywords like `design' or `pattern' along with some other related words, the text can be easily classified into the design class. However, relying on some specific words and classifying a long text based on some discrete keywords often can be misleading. We first analyze the top 100 words in each class (design/not-design) and analyze the overlap between the words. An overlap means that a particular word is present in texts that relate to both of the classes and it falls into one of their top 100 words. This overlap implies that word would be a poor indicator of class membership.


\begin{table}[!ht]
\centering
\caption{The percentage of overlap in top 100 words and top 100 tri-gram phrases---words/phrases occurring in both design and general (not design) wordlists.}
\label{tab:overlap}
\begin{tabular}{cc}
\toprule
n-gram features & \% of overlap  \\ \midrule
Top 100 words & 46.0  \\ 
Top 100 phrases & 9.3\\ \bottomrule
\end{tabular}
\end{table}

The first data row of Table \ref{tab:overlap} shows a 46\% overlap of top words in the design and general class. This suggests that relying on a particular set of discrete keywords can often lead to misclassification since both classes contain similar kinds of keywords. To deal with this overlap, we instead rely on a tri-gram model. A tri-gram model contains three words each time. This way, the middle word can have the context of both the neighbouring words surrounding it. The second data row of Table \ref{tab:overlap}) illustrates a significant reduction in the amount of overlap between the top tri-gram phrases. This way, the classes can have features that are very unique to that particular class. Although it is possible to reduce the overlap by increasing the size of the phrases, we opted for using a tri-gram model because of its proven capability of being the best predictor for probabilistic measures \cite{tremblay2011}.  

%% file: sections/04-classifying-design-discussion.tex
\subsection{Resolving Potential Transferability Issues}
\label{classifying-design-discussion}
The previous challenge was about expanding the set of software specific data we can use to vectorize and train our model.
Our final two challenges refer to building software context into the model, both for context based on data domain (issue/comment/pull request) and based on software semantics, rather than general language models, which are typically trained on Wikipedia or other general purpose sources. To resolve these problems, we introduce data augmentation and software word vectors.

\subsubsection{Data Augmentation Using Similar Word Injection}
\label{subsec:swi}
Machine learning models typically require a large amount of data to train, test and validate. The previous studies that we explored in Section \ref{sect:related_work} either reported manual labeling or synthetic generation of the data. 
We used a relatively large amount of data containing 260,000 rows of discussion from \SO{} for this study, yet, this is only a very small subset of the total discussions in \SO{}. Thus, augmenting this data might help with performance.

\begin{figure}[htbp]
    \centering
    \includegraphics[width=\textwidth]{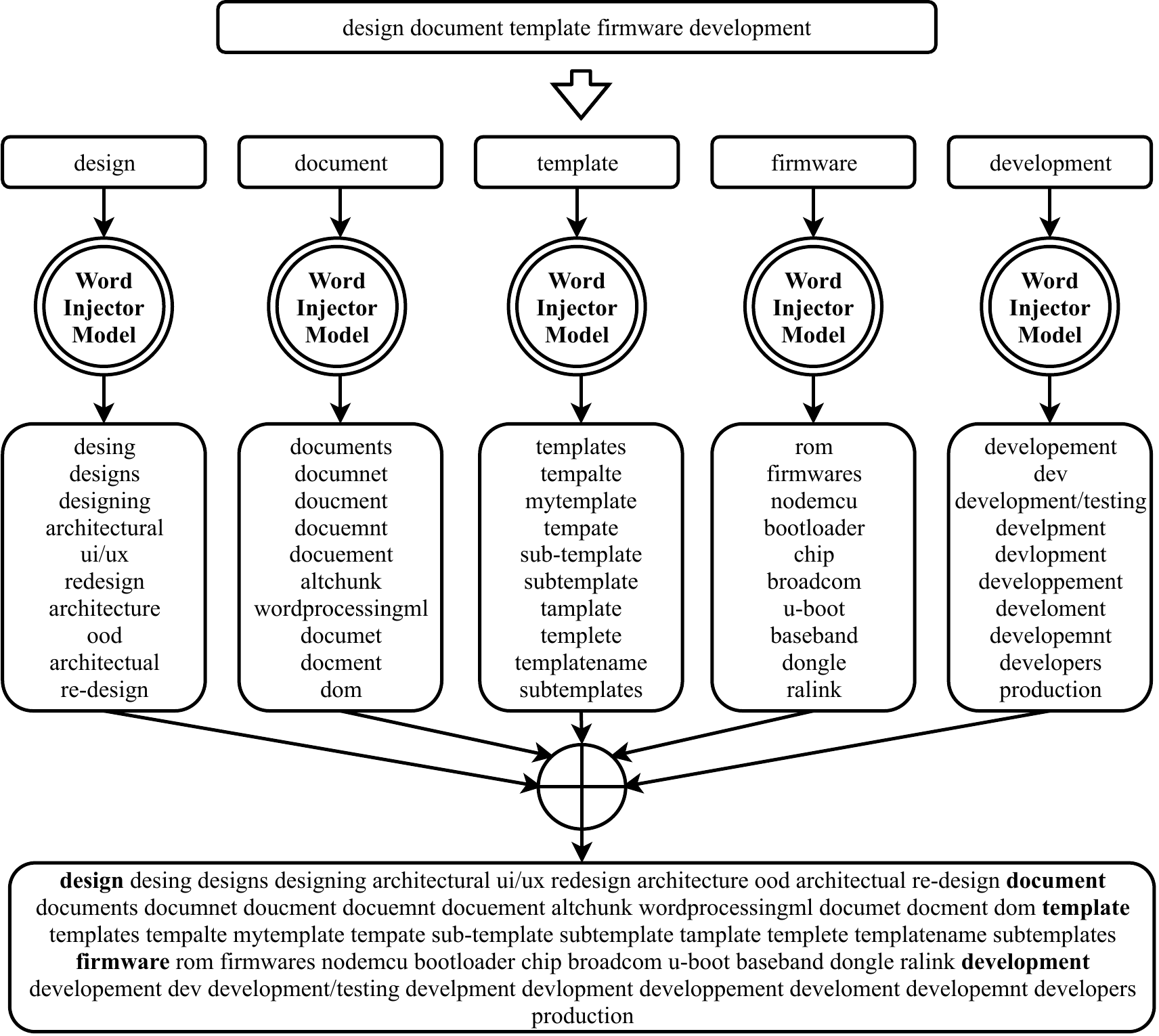}
    \caption{Similar word injection workflow}
    \label{fig:similar_word_injection}
\end{figure}

To augment the data we collected, we use an unsupervised word embedding from the study of Efstathiou, Chatzilenas, and Spinellis \cite{Efstathiou2018}  (which we call ECS) to inject similar words into the small subset of data we can have. The ECS word embedding was trained on 15 Gb of text obtained from \SO{}. This unsupervised model, as the name suggests, only needs data but does not require any labeling. On the other hand, the similar word injection into the small subset of data enables the data to have all the possible vocabulary from the domain with the surrounding context. We call this approach ``providing total domain context''. Fig \ref{fig:similar_word_injection} shows an example: the word `design' is augmented with, among others, `redesign' and `architecture'. 

 Our injection algorithm first splits the training corpus into individual words. Then, each word is passed through the model to obtain a set of similar words for that specific word. The Efstathiou model outputs a similarity index from 0 to 1 of each word in the set of similar words.  We take those words which have a similarity index more than or equal to 0.6, after experimenting with different cutoff numbers to find an optimal AUC result. Then, we concatenate the set of similar words with the actual word to obtain the revised (augmented) corpus.

\subsubsection{Software Specific Word Vectorizer}
\label{subsec:swv}
Since we are using an embedding model to turn our text data into a numeric vector, it is important that our word embedding model has the context of software engineering since we are dealing with texts that are mostly about software. In our previous study, we used the GloVe word embedding \cite{pennington2014glove} which is trained on Wikipedia data. This word embedding works well for general purpose text classification, however, it does not perform well with text from from software domain. For example, the word `class' has general contextual relation with education while in software engineering, the word `class' is an integral vocabulary of object oriented design. To address this issue, we collect our data for training the word embedding model from literature related to software engineering (described in \S \ref{subsec:word_embedding_dataset}).

\begin{figure}[htbp]
    \centering
    \includegraphics[width=\textwidth]{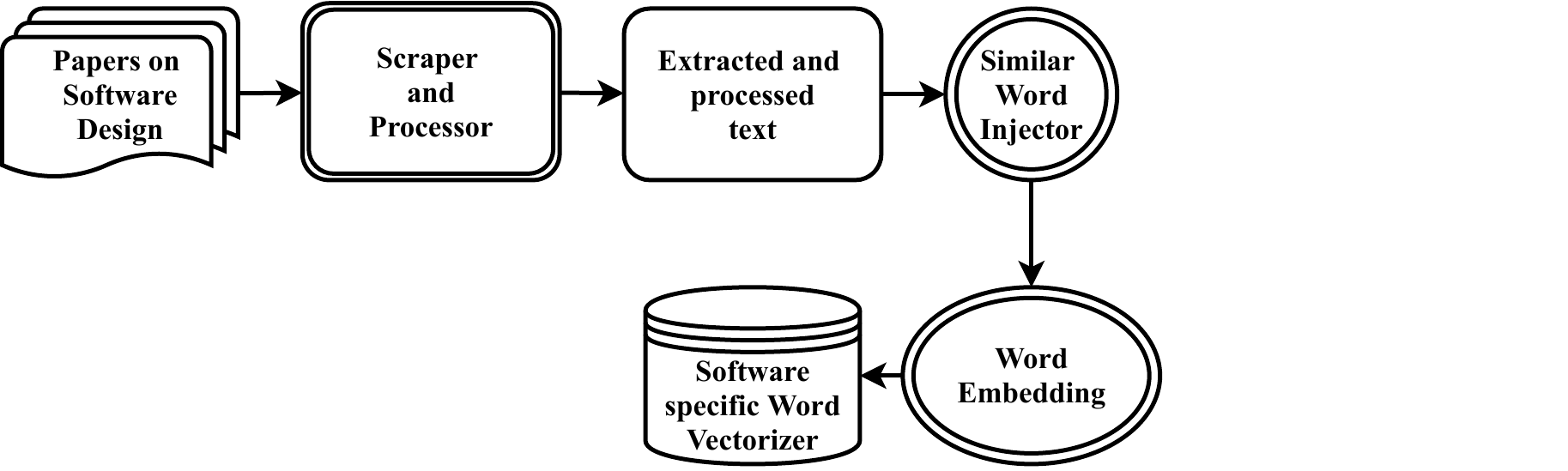}
    \caption{Training the Word Vectorizer model}
    \label{fig:word_vectorizer}
\end{figure}

Joulin et al. \cite{joulin2016} described the use of subword information to enrich a word vector. They also used a similar technique to implement a compression algorithm \cite{joulin2016a} for classification models. We implement the algorithm by Bojanowski et al., \cite{bojanowski2016} with the help of fasttext\footnote{https://fasttext.cc/}  with the steps illustrated in Figure \ref{fig:word_vectorizer}. The software literature corpus is used to train our new word embedding model. 

First, the dataset is passed through the processing steps described in section \ref{subsec:data_processing}. Then, each word of the corpus is evaluated and injected with similar words using the ECS similar word injector. During the training phase, we train the classifier unsupervised since we just want to group the data according to similarity. We have used skipgram \cite{mikolov2013} as one study \cite{mikolov2013skipgram_improve} shows that skipgram models works better with subword information than cbow \cite{mikolov2013}. We take words with length from 4-20. Since we are removing every word less than three characters in our text processing step, it is not important to take the words less than 4 characters. In addition, design words seem to be on the longer side, for example `reproducibility' contains 16 characters. We are considering characters up to 25 characters in length. We take 300 dimensions of each word training by looping for 10 epochs. Both of our decisions of taking 300 dimension and looping for 10 epochs is taken because the training corpus is relatively small.

\subsubsection{Providing and Transferring Context}
\label{subsec:context_transfer}
In our previous study \cite{Mahadi_2020} we found that both the lack of domain context in a small subset of the total data, and making the train data and test data similar in terms of vocabulary, are a challenge. The out-of-vocabulary problem is one such instance, since word vectors are not helpful if they do not contain a particular word (such as a unique source code identifier) that is out of its vocabulary. 

Appropriate classification of text requires the training data to have proper understanding of the whole domain, e.g., of projects and artifact types. Ideally, the test and train data also have similar context. For example, training on sentences from pull requests from Project A will have poor results if applied to code comments from Project B. Nonetheless, we expect a design classifier to be able to understand (as humans would) that both are valid, and frequent, ways in which design discussions happen. 

Vocabulary from one domain (ex. \SO{}) can differ from another communication medium (e.g., Github, email). For example, most of the discussion happening in \SO{} relates to questions and answers, while Github represents mostly statements in the form of issue tracking and pull request. Hence, the vocabulary and the context also varies from one domain to another. 

We create an unsupervised word vector trained on data with a similar context to the test domain (in this case, data from Github) to inject similar words into the training data (\SO{} data). 
We also repeat this step in the opposite direction (reusing the already trained ECS similar word injector model) to inject the training domain's vocabulary context to the test data. We name this step  ``cross domain context transfer''. 

\begin{figure}[htbp]
    \centering
    \includegraphics[width=0.7\textwidth]{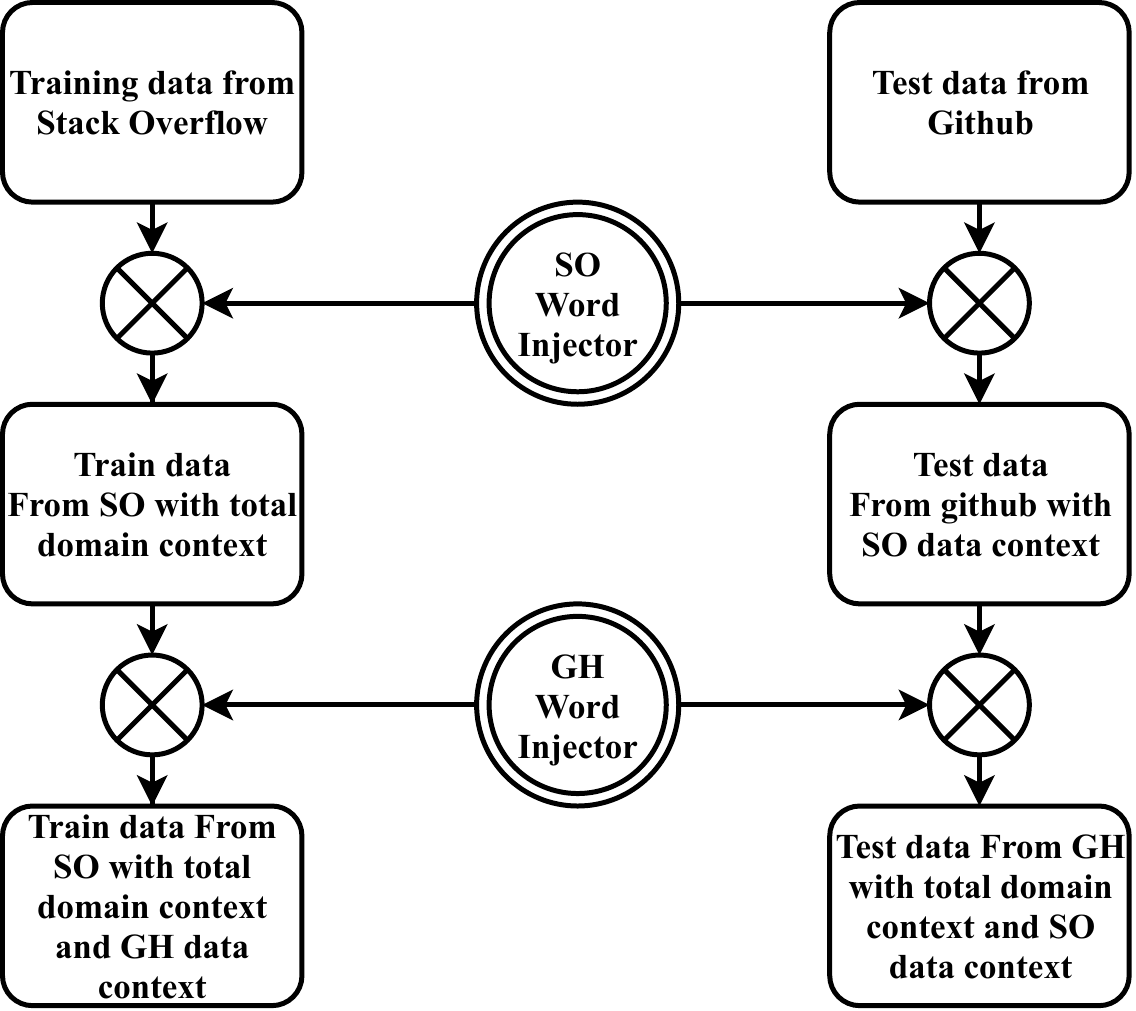}
    \caption{Proposed idea of providing total and cross domain context}
    \label{fig:context_transfer}
\end{figure}

In transferring context, we are biasing the models to the software engineering context, but we must be careful not to further bias the training data with information from the test data set. We carefully and completely separate the test and train data from the training of the word injector models.
The Github word vector is trained on a random selection of pull request text obtained from the Github BigQuery data dump. However, it is possible that a few of the 1000 items (pull request discussions) from the Brunet2014 dataset appear in our random selection in the Github word vector, and it is possible that some words in the 260,000 sentences we use from \SO{} also appear in the ECS word vector (trained on a random selection of 15 Gb of \SO{} text). However, we judged this risk to be minimal, given the data volumes involved, and the limited impact that a single significant word can have on the training outcome. We outline the context transfer approach in Fig. \ref{fig:context_transfer}.

\section{Study Design}
\label{sec:design}
\begin{figure}[htbp]
    \centering
    \includegraphics[width=\textheight,angle=-90]{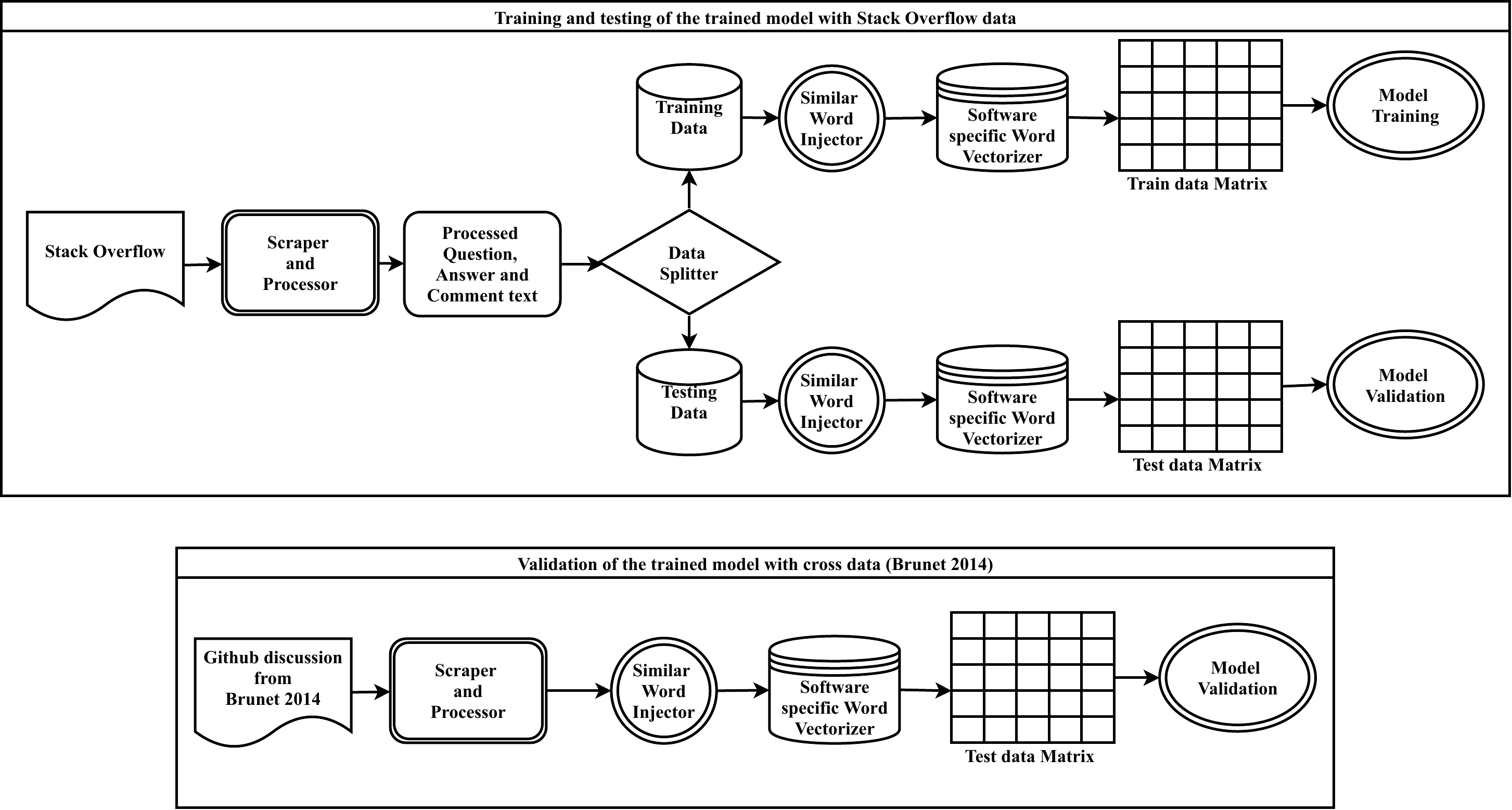}
    \caption{High level design of the study}
    \label{fig:study_design}
\end{figure}

In this section, we describe the overall design of our study along with the proposed design of each component of our system architecture. Figure \ref{fig:study_design} illustrates an overall and high level view of our study with all the components. We have talked about the source of the data and how to acquire and process the data in Section \ref{subsec:classifier_dataset}. After processing, the data goes through several models and is transformed into a matrix. Similarly, the classes are also expressed as a vector. We refer to the train data  matrix and train class vector as train vector. Similarly we name the combination of the text matrix and its associated vector, test vector. 

We use 10 classifiers: `Nearest Neighbors', `Decision Tree', `Random Forest', `Logistic Regression', `Gaussian Naive Bayes', `Neural Net', `AdaBoost', `QDA', `Linear SVM', `RBF SVM' and analyzed the output of the model with every transformation of the data, using Scikit-learn \cite{scikit-learn}. We start by implementing the similar word injector model to inject similar vocabulary into the data. Then we use our software specific word vectorizer to turn the text into matrices and vectors. 

We have used a Google Cloud Platform\footnote{https://cloud.google.com/} instance with processor of Intel's E2 platform and 16 GB memory to load four models (1 word vectorizer, 2 word injectors, 1 classifier) at once.

We split the total dataset of 260,000 rows into three parts\footnote{for access, see \replicationUrl{}}. We used 200,000 rows of data for training the model while keeping 30,000 rows as validation data and the other 30,000 rows as within-dataset test data (we still use the Brunet2014 dataset for the cross-domain problem). We use our validation data in training the neural networks to validate the results after each iteration. Our test data set is kept completely separate from the training phase. Hence, our test data can be considered to be an unknown set of data \textbf{from the same domain}. After the completion of the training, we evaluate our models with the test data. 

%% file: sections/05-results-discussion.tex
\section{Experiments and Results}
\label{results-discussion}
\begin{figure}[h]
    \centering
    \includegraphics[width=\textwidth]{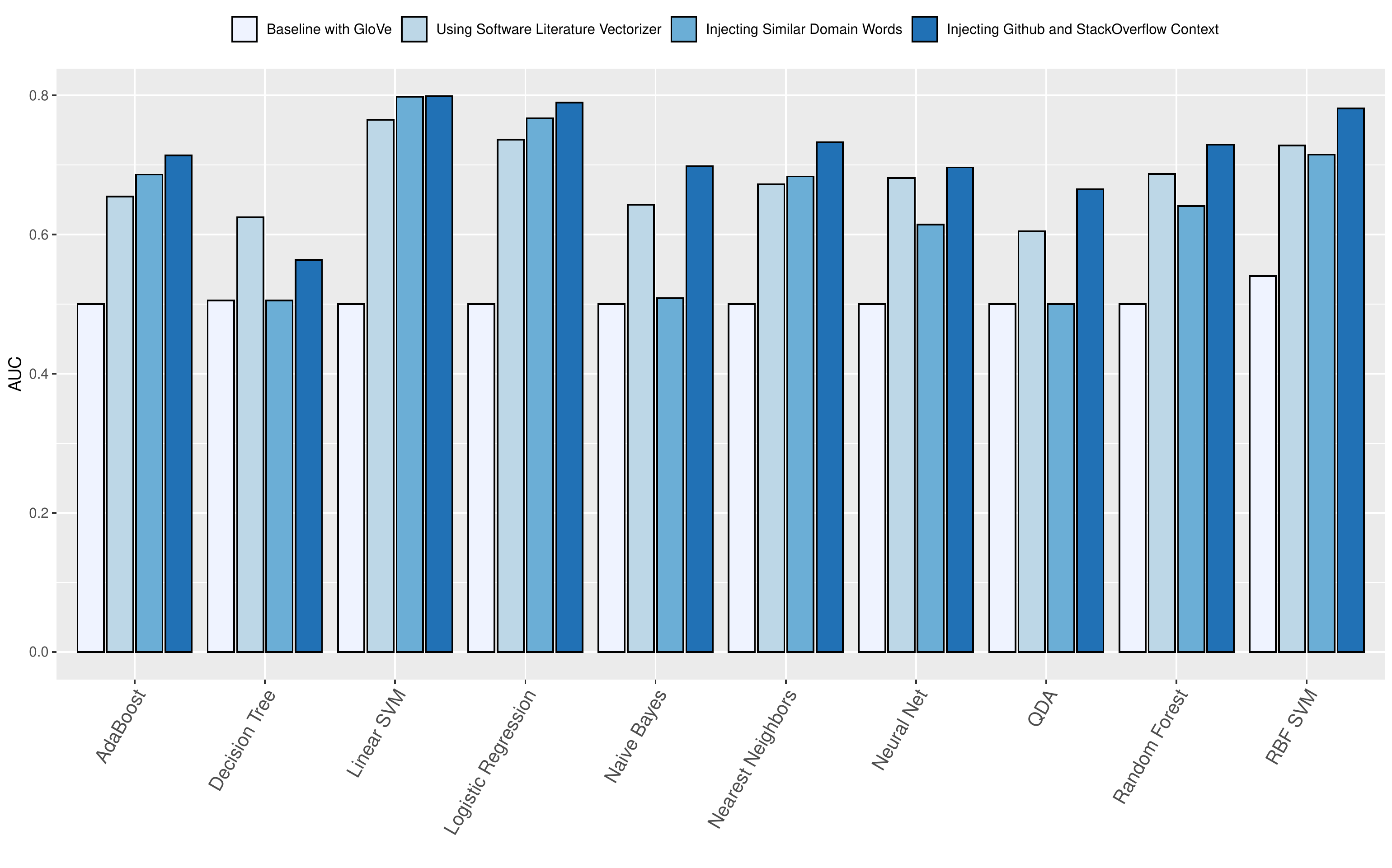}
    \caption{Experimental results. Labels reflect AUC.}
    \label{fig:all_results}
\end{figure}

Our main goal of this study is not simply to show a new state of the art (SOTA) result outperforming a previously studied classifier when classifying data from the \emph{same} domain. Rather, we want to illustrate how the data can be generalized and fed to the model so that the model can perform better than the previous studies in terms of detecting design discussions from an unknown domain. We used 10 modern classifiers to demonstrate the performance of each classifier individually. We report on the AUC score of all classifiers in Figure \ref{fig:all_results}. Each bar in a classifier group in that figure reflects either the baseline or one of the experiments we describe below. Table \ref{tab:result-table} shows a summary of those same results.
 We conduct our experiment in two stages. We first look at how using software word vectors improves the results. Next, we consider data augmentation using similar words. Results reflect the AUC score of a classifier trained on the 200,000 rows of \SO{} discussions and tested on the Brunet2014 data (Github pull requests).

\input{all3_auc.tex}

\subsection{Experiment 1: Software Literature Word Vector}
\label{subsec:software_specific_we_result}
First, we want to make sure that our hypothesis of domain specific word vectorizer performs better than the conventional word embedding model to answer RQ 2. For this purpose, we compare the performance of our software literature word vectorizer with a state-of-the-art word vectorizer, GloVe.  Except for using two different word vectorizers, all the other constraints were kept same during the two test runs.

The left 2 bars in Fig. \ref{fig:all_results} illustrates the comparison of our vectorizer with GloVe vectorizer. The horizontal axis represents the classifiers used and the vertical axis shows the performance of different classifiers in terms of AUC. Using GloVe as the vectorization model provides AUC scores in the range of 0.5 - 0.55. The performance is significantly improved while vectorized with our software specific word embedding model as shown with the second bar in each group. Thus we conclude that the software specificity of our new vectorization (trained on 300 literature sources from the SE domain) greatly improves classification results. What seems to be happening is that the vector space represented by the software specific taxonomy is better able to distinguish design-related words from general software words, unlike the Wikipedia-based GloVe. The best classifier in this experiment was Linear SVM, with AUC of 0.765 (Precision: 0.695, Recall: 0.672, MCC: 0.492).

\subsection{Experiment 2 and 3: Data Augmentation Results}
\label{subsec:data_augmentation_result}
After success on the first stage, we augment the dataset by injecting similar words into the data according to Figure \ref{fig:context_transfer}. First, we explore how similar word injection performs in providing \emph{total domain context}. 

The performance after augmenting (only) the training dataset with software literature words is shown in the third bar of each classifier group ("Using Software Literature Vectorizer") of Figure \ref{fig:all_results} and column 3 of Table \ref{tab:result-table}. As seen from the figure, we get a mixed range of performance by this approach. The performance of most of the classifiers decreases or show only marginal increases over the baseline (column 2) following this method. This is because the injection of similar words only to the training data increases the statistical weight and bias towards only one domain (namely \SO{}). 

On the other hand, because the test data is from another domain (Github), the high bias in one domain prevails during the probability analysis resulting in missed classification of the test data. The best classifier in this experiment was Linear SVM, with AUC of 0.798 (Precision: 0.647, Recall: 0.901, MCC: 0.581).

%

This problem of high bias is removed by using instead a cross domain similar word injection, the rightmost bar in Figure \ref{fig:all_results} and the right column of Table \ref{tab:result-table}.
Results shows that AUC score of all classifiers is higher after augmentation using cross domain similar word injection. Some of the scores, namely Linear SVM at 0.80 (Precision: 0.734, Recall: 0.771, MCC: 0.592), are SOTA in \emph{cross-domain} design discussion classification studies. The use of the contextual words from both domains, essentially transferring the characteristics of discussions in those domains to either dataset, has improved classification performance.

Most of the related work in classifying design discussion does not report or apply cross domain classification. Only one study by Viviani et al. \cite{Viviani2019} attempts to show the performance of their classification while classifying an unknown dataset. A gold standard dataset was manually created by them which was used to justify the performance of their model with unknown data. While that classifier produced better results than ours, the gold standard dataset was not cross domain data. It was made from Github discussions similar to the training data. We believe that this study is unique in using dataset, vectorizers and attempts to tackle cross domain design discussion classification. 

Precision and recall measures show the classifier is also balanced between false positives (low precision) and false negatives (low recall). For the context of using design classification to augment developer activity, these values are reasonable for indicating whether a pull request (for example) is design-related or not (roughly 27\% of the time it will be mischaracterized), or in retrieving all the design-related discussions in a search (which will miss 23\% of the discussions). More investigation with human stakeholders is necessary to explore how these values---and also which discussions are missing---impact development activities.

\begin{figure}
    \centering
    \includegraphics[width=0.8\textwidth]{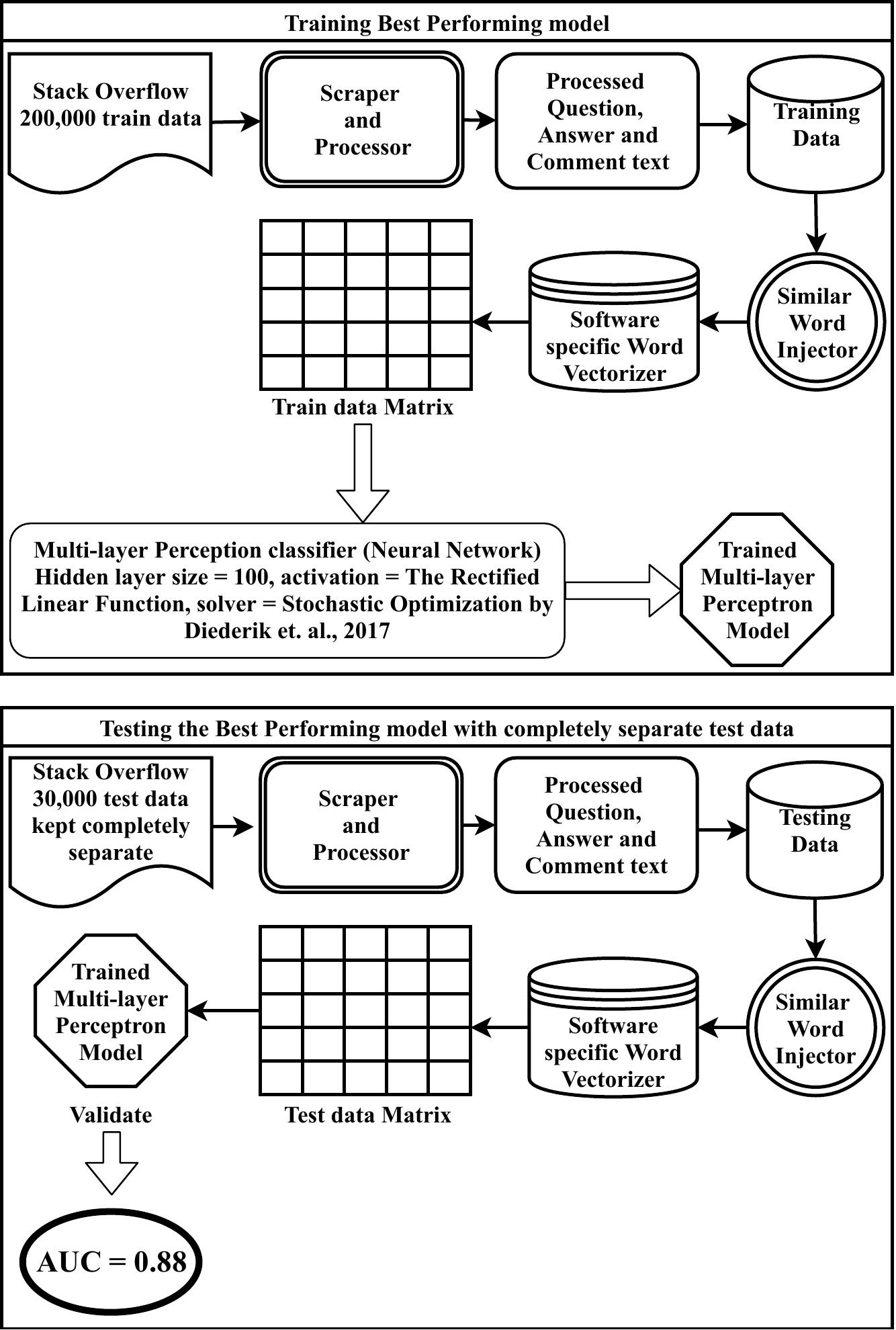}
    \caption{Protocol map of the test data validation}
    \label{fig:test-data-validation}
\end{figure}

\paragraph{Within Dataset Results}
We use the software specific word vectorizer along with similar word augmentation for train and test data and a simple neural network (Multi-layer Perceptron classifier) illustrated in Figure \ref{fig:test-data-validation} yielding 0.88 AUC (MCC: 0.76), which is a state-of-the-art score for classifying unknown data within the \emph{same} domain (in this case, \SO{} design discussions). This compares directly with the approach \textsf{NewBest} we described in \S \ref{sect:extending_the_replication} which obtained AUC of 0.84 (MCC 0.63). Note that this is expected, since there is no cross-domain data differences (as there was with the Brunet2014 data).

In summary, we have demonstrated how we might overcome three challenges in design mining studies. We show substantial improvements in classification accuracy across domains and projects by using more labeled data, software specific context, and cross-domain context. We also present a state-of-the-art approach for classifying unknown discussion within the domain.

%% file: all3_auc.tex
\begin{table}[ht]
\caption{Tabular form of Figure \ref{fig:all_results}. Baseline is the GloVe vectorizer; Literature the experiment using the software literature vectorizer; Domain Words the injection of software domain words into the training data; and Cross Context the injection of context information from Github and StackOverflow.}
\label{tab:result-table}

\centering
\begin{tabular}{rllll}
  \toprule
Experiment & Baseline & Literature & Domain Words & Cross Context \\ \midrule
  Nearest Neighbors & 0.5000 & 0.6720 & 0.6833 & 0.7324 \\ 
  Decision Tree & \textbf{0.5051} & 0.6245 & 0.5052 & 0.5633 \\ 
  Random Forest & 0.5000 & 0.6867 & 0.6406 & 0.7288 \\ 
  Logistic Regression & 0.5000 & 0.7364 & 0.7671 & 0.7894 \\ 
  Naive Bayes & 0.5000 & 0.6424 & 0.5082 & 0.6978 \\ 
  Neural Net & 0.5000 & 0.6810 & 0.6140 & 0.6963 \\ 
  AdaBoost & 0.5000 & 0.6542 & 0.6859 & 0.7134 \\ 
  QDA & 0.5000 & 0.6046 & 0.5000 & 0.6650 \\ 
  Linear SVM & 0.5000 &\textbf{0.7648} & \textbf{0.7978} & \textbf{0.7985} \\ 
  RBF SVM & 0.5400 & 0.7280 & 0.7144 & 0.7812 \\ \midrule
  Mean & 0.5045 & 0.6795 & 0.6417 & 0.7166 \\
   \bottomrule
\end{tabular}
\end{table}

%% file: sections/06-discussion.tex
\section{Discussion}
\label{discussion}

We discuss the implications of our results for future design mining studies, how dataset size matters, discuss the ways to improve future studies and account for researcher degrees of freedom. We begin with threats to validity for this work.

\subsection{Limitations: Bad Analytics Smells}
We use the concept of bad analytics smells from Menzies and Sheppherd \cite{Menzies2019}. In that paper, the authors introduce a succinct list of twelve potential study design flaws in analytics research, and suggest some mitigations. Here, we list the smells this paper might emit, and ignore the ones we believe we have dealt with or do not apply. 

\begin{enumerate}[label=(\alph*)]
  \item \emph{Using suspect data}: we rely extensively on the tags that the author and moderators of \SO{} created. However, we conduct some statistical test to validate the data labels. We also re-use previous datasets, but several different ones, and contribute a new dataset to the literature. As we mentioned, it is possible the context transfer approach tainted the test data with training context (and vice versa), since we rely on word vectors from the same domains as a fine-tuning approach. However, the specificity of the data involved and the overall large volumes of data used to train the vectors makes us think this risk is minimal.
  \item \emph{Low power}: ultimately, the design mining data relies on a limited set of labeled data (or makes the possibly invalid assumption that the tagging in \SO{} reflects real design). Below, we show some results that suggest we do indeed have sufficient training data, or rather, that bigger increases in performance will need to come from somewhere other than more data.
  \item \emph{No data visualizations}: we present a limited set of visualizations because text data is always difficult to visualize. However, we provided some statistical analysis in the data validation in Section \ref{subsec:data_validation}.
  \item \emph{Lack of manual validation}: in this study, we mainly wanted to explore the improvements of having a large and related data on the classification results. We present very little manual validation of the data we chose to have during our data collection phrase. We decided to go for the tags we felt to be closer to design topics. However, there was not any manual validation done on the data we get from the tags. The classification result should be improved with a careful choice of tags and data that is clearly related to the design topics.
  \item \emph{Not tuning}: Because of our focus on making the vectorizer and data generalized for models, we did not emphasize optimization for any specific model. In the case of neural network, we use the most simple one with default optimization. Thus, it is possible our results are an under-estimate of a properly tuned model. 
We use a cutoff point of 0.6 for word similarity index in data augmentation. We experimented with several similarity index and 0.6 seem to work the best in terms of AUC and the size of vocabulary. However, we do not validate this choice during our experiments of this paper. One other limitation is that we used 300 different literature sources to extract text to train our word vectorizer. The basis for selecting these 300 was a random one, based on common SE literature. We took this approach to remove bias towards a specific kind of data. However, this was a missed opportunity for us to direct our model in a very specific way (e.g., by only mining design related papers or texts). Furthermore, the disadvantage of using a limited number of literature sources is reflected in the total number of vocabulary in the training corpus of the vectorizer.
  \item \emph{Not justifying choice of learner}: we report on the standard machine learning approaches including some deeper network models. It is possible that some other learner could improve results, particularly a tuned deep model.
  \item \emph{Not exploring simplicity}: this smell argues that we should ensure we do not overfit our results with complex models. We explicitly test our models on non-similar (cross-project/domain) datasets. 
\end{enumerate}

\subsection{Choice of Training Size}
\begin{figure}
    \centering
    \includegraphics[width=.7\textwidth]{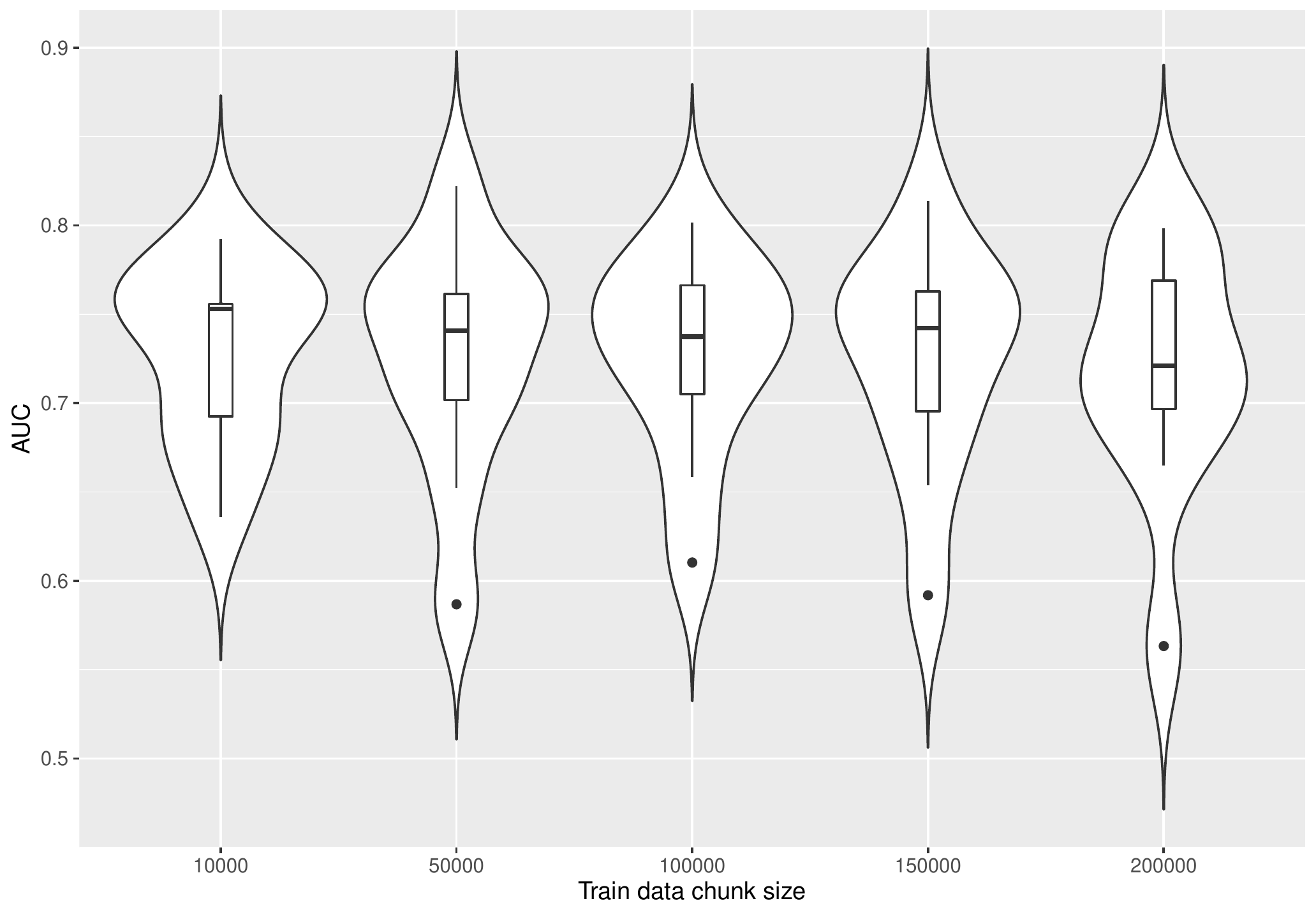}
    \caption{Performance of the 10 classifiers after training with different size of train data. Dot represents an outlier, solid bar is the median.}
    \label{fig:box_plot}
\end{figure}

The violin plot at Figure \ref{fig:box_plot} represents the distribution of the AUC of the 10 classifiers with different size of chunks of the training data. We ran all of the preceding experiments using a chunksize of 200,000 data points, a figure selected based on the data availability. However, we were curious if less data might return a reasonable performance as well. We experimented with four chunks: 10,000,  50,000, 100,000, 150,000, 200,000. The x-axis represents the chunk size. From the plot, we can see that chunk of 10,000 train data has the highest median (dark black line inside the boxes) with relatively high range for minimum and maximum. 

This implies this chunksize may well outperform datasets 10x larger. While this seems counter-intuitive, the issue of dataset size in machine learning is still an active area of research \cite{Sun2017}. For example, in deep learning models larger datasets seem to make a bigger difference. We use less complex models such as SVM, and thus our data---for example, from Stack Overflow---may be saturated for that algorithm more quickly. More research into the learning curve (error vs accuracy) for design mining is necessary. This choice is another example of researcher degrees of freedom that should be answered early in the study. It also seems to contradict the first challenge we noted in \S \ref{dataset-design-discussion}, that dataset size was important. Our belief is that this is still true, but that we are seeing diminishing returns after we increase by an order of magnitude more data (e.g., Brunet2014 was 1000 records, roughly 20\% of which are design, vs our small sized \SO{} dataset of 10,000, where roughly 5,000 are design).


\subsection{The Effectiveness of Software Specific Vocabularies}
Most neural language models are intended for general-purpose language tasks, such as translation and question answering. However, these models do not deal well with domain-specific terminology, because such terminology is a small fraction of the overall training data \cite{Efstathiou2018}.

Introducing software vocabularies resulted in a big improvement in accuracy of the design mining classifier, suggesting that even fairly simple contextual approaches, such as similar word augmentation, and software specific vectors, can be a big help.  The importance of individual project context, as opposed to general software context, is still unclear (and training models is less useful with less data to work with). Novielli et al. suggest that fine-tuning for sentiment analysis, at the project level, can be quite helpful \cite{novielli20}. We intend to investigate how these vector models can help with fine-tuning general purpose models like ULMFiT or BERT, as suggested (for source code) by recent work from Karampatsis et al. \cite{karampatsis20}.

\subsection{Researcher Degrees of Freedom}
As our protocol maps have shown (cf. Fig. \ref{fig:brunet-protocol}, Fig. \ref{fig:context_transfer}), these studies have many possible analysis paths: datasets, training algorithms, test/train data, vectorization, stop words, etc. These researcher degrees of freedom mean there are many ways to achieve results in these studies, and picking a single point result as the SOTA may be misleading. At the extreme researcher bias \cite{romano20} may be involved, typically implicitly.

We think there are three major steps to take to help solve the RDOF question, and improve conclusion stability.

\begin{enumerate}
  \item Use protocol maps or other graphical models to clearly outline the degrees of freedom, and chosen paths. Improve study reporting in general. There are lessons to be learned from scientific workflow software already well-developed in, for example, high-energy physics. This makes replication of results simpler and direct comparison possible.
  \item For confirmatory studies, pre-registered hypotheses and protocols, like in medicine, make it clear what conclusions are valid, and which might be conditioned on the observed results.
  \item Improve understanding of the concept of RDOF. Develop tools that can automatically generate protocol maps based on common data science pathways. For example, we could apply our existing strengths in software slicing to analyze available parameter choices and dependencies in machine learning frameworks.
\end{enumerate}

%% file: sections/07-conclusion.tex
\section{Conclusion}
\label{conclusion}

Like Shakiba et al. \cite{Shakiba2016} and Viviani \cite{Viviani2018b,Viviani2019}, we envision a design tagging tool that can be applied broadly to all design discussions. This would be a necessary first step in automatically analyzing design decisions and recommending alternatives or improvements. To get to that point, the community needs to increase the amount of data available for these sorts of mining tasks. \SO{}, as we demonstrate, is one potentially rich source for labeled data. More importantly, a better understanding of the nature of design discussions is needed. Expanding on qualitative studies such as Viviani et al. \cite{Viviani2018b} or those surveyed in van Vliet and Tang \cite{vanVliet2016} are the likely way forward, now that others, such as Brunet et al. \cite{Brunet2014} have shown the nature of specific design mining challenges within a constrained context. 

The importance of design mining to practitioners is largely speculative at this point, as researchers try to improve state of the art (SOTA) to the point it can be useful in practice. More effort in bringing even preliminary design mining results into prototype tools is important to understand how (and why) practitioners might use design mining. 

This study represents a continuation of our previous work \cite{Mahadi_2020} by introducing a concept of building software specific word vectorizer to improve word embedding for software engineering related discussions backed by results on applying it on a wide variety of classifiers. We showed state of the art results on the cross project classification problem using this approach. This paper also demonstrates the application of augmentation (similar word injection) to transfer context between cross-domain with experimented results and discussions. We also restrict ourselves from going further with the classifiers by tuning the hyper parameters and exploring plausible explanation on the difference in performance for the different classifiers. However, we believe that our idea of software specific word embedding and context transfer, along with efficient choice of classifier and parameter optimization, can improve the study of design discussion mining even further.

%% file: sections/related_table.tex
\setcounter{table}{0}
\renewcommand{\thetable}{A\arabic{table}}
\begin{landscape}
\begin{longtable}{cp{1.4cm}p{1.8cm}p{1cm}p{1.8cm}p{1.8cm}p{3cm}p{2cm}}
\caption{Comparison of recent approaches to design discussion detection. Effectiveness captures the metric the paper reports for classifier effectiveness (accuracy, AUC, precision, recall, F1). A * indicates the metric is calculated on imbalanced data. NB: Naive Bayes; LR: Logistic Regression; DT: Decision Tree; RF: Random Forest; SVM: Support Vector Machine}
\label{table:related}\\
\textbf{Study} & \textbf{Projects Studied} & \textbf{Data Size} & \textbf{ML Algorithm} & \textbf{Effectiveness} & \textbf{Prevalence} & \textbf{Defn. of Design} & \textbf{Defn. of Discussion} \\
\endfirsthead
\multicolumn{8}{c}%
{{\bfseries Table \thetable\ continued from previous page}} \\
\textbf{Study} & \textbf{Projects Studied} & \textbf{Data Size} & \textbf{ML Algorithm} & \textbf{Effectiveness} & \textbf{Prevalence} & \textbf{Defn. of Design} & \textbf{Defn. of Discussion} \\
\endhead
Brunet \cite{Brunet2014} & 77 high importance Github projects & 102,122 comments & NB DT & \textbf{Acc*: 0.86/0.94} & 25\% of discussions & Design is the process of discussing the structure of the code to organize abstractions and their relationships. & A set of comments on pull requests, commits, or issues \\
\vspace{1em}
Alkadhi17 \cite{Alkadhi2017} & 3 teams of undergrads & 8,702 chat messages of three development teams & NB SVM + undersampling & \textbf{F1: 0.85} & 9\% of messages & Rationale captures the reasons behind decisions. & Messages in Atlassian HipChat \\
Alkadhi18 \cite{Alkadhi2018} & 3 Github IRC logs & 7500 labeled IRC messages & NB SVM & \textbf{F1: 0.77} & 25\% of subset labeled & Rationale captures the reasons behind decisions. & IRC logs from Mozilla \\
Zanaty \cite{Zanaty2018} & OpenStack Nova and Neutron & 2817 comments from 220 discussions & NB  SVM KNN DT & \textbf{AUC: 0.85} & 9-14 & Brunet's \cite{Brunet2014} & Comments on code review discussions \\
Shakiba \cite{Shakiba2016} & 5 random Github/SF & 2000 commits & DT RF NV  KNN & \textbf{AUC: 0.53} & 14\% of commits & None. & Commit comments \\
Motta \cite{Motta2018} & KDELibs & 42117 commits, 232 arch & Wordbag matching & N/A & 0.6\% of commits & Arch keywords from survey of experts & Commit comments \\
Maldonado \cite{Maldonado2017} & 10 OSS Projects & 62,566 comments & Max Entropy & \textbf{F1*: 0.403 } & 4\% design debt & Design Debt: comments indicate that there is a problem with the design of the code & Code \\
Viviani18 \cite{Viviani2018} & Node, Rust, Rails & 2378 design-related paragraphs & N/A (qualitative) & N/A & 22\% of paragraphs & A piece of a discussion relating to a decision about a software system’s design that a software development team needs to make & Paragraph, inside a comment in a PR \\
Viviani19 \cite{Viviani2019} & Node, Rust, Rails & 10,790 paragraphs from 34 pull requests & RF & \textbf{AUC 0.87} & 10.5\% of paragraphs & Same as Viviani18 & Same as Viviani18 \\
Arya19 \cite{Arya2019} & 3 ML libraries & 4656 closed issue sentences & RF & F1: 0.69 & 30\% of sentences & ``\emph{Solution Discussion} ... in which participants discuss design ideas and implementation details, as well as suggestions, constraints, challenges, and useful references ''.  & A closed issue thread \\
Mahadi20 \cite{Mahadi_2020} & Stack Overflow discussions & 51,990 questions and answers & LR/ SVM/ ULMFiT & \textbf{AUC: 0.84} & N/A & A question or answer with the tag ``design" & \SO{} question/answer \\
(This paper) & \SO{} discussions and literature & 260,000 questions, answers and comments & multiple, including SVM & \textbf{AUC: 0.80 (cross-domain)} & N/A & \SO{} questions with tags related to ``design" & \SO{} question/ answer/ comments \\
 \midrule
Meta & Open-source & 46,470 & N/A &\textbf{AUC: 0.80-0.87} & 15.25\% & N/A & N/A \\ 
\bottomrule

\end{longtable}
\end{landscape}

%% file: emse.bbl
\begin{thebibliography}{10}
\providecommand{\url}[1]{#1}
\csname url@samestyle\endcsname
\providecommand{\newblock}{\relax}
\providecommand{\bibinfo}[2]{#2}
\providecommand{\BIBentrySTDinterwordspacing}{\spaceskip=0pt\relax}
\providecommand{\BIBentryALTinterwordstretchfactor}{4}
\providecommand{\BIBentryALTinterwordspacing}{\spaceskip=\fontdimen2\font plus
\BIBentryALTinterwordstretchfactor\fontdimen3\font minus
  \fontdimen4\font\relax}
\providecommand{\BIBforeignlanguage}[2]{{%
\expandafter\ifx\csname l@#1\endcsname\relax
\typeout{** WARNING: IEEEtran.bst: No hyphenation pattern has been}%
\typeout{** loaded for the language `#1'. Using the pattern for}%
\typeout{** the default language instead.}%
\else
\language=\csname l@#1\endcsname
\fi
#2}}
\providecommand{\BIBdecl}{\relax}
\BIBdecl

\bibitem{Gousios2014}
\BIBentryALTinterwordspacing
G.~Gousios, M.~Pinzger, and A.~v. Deursen, ``An exploratory study of the
  pull-based software development model,'' in \emph{Proceedings of the 36th
  International Conference on Software Engineering}, ser. ICSE 2014.\hskip 1em
  plus 0.5em minus 0.4em\relax New York, NY, USA: Association for Computing
  Machinery, 2014, p. 345–355. [Online]. Available:
  \url{https://doi.org/10.1145/2568225.2568260}
\BIBentrySTDinterwordspacing

\bibitem{Kazman2016}
R.~Kazman and H.~Cervantes, \emph{Designing Software Architectures: A Practical
  Approach}, ser. SEI Series in Software Engineering.\hskip 1em plus 0.5em
  minus 0.4em\relax Addison-Wesley, 2016.

\bibitem{woods16}
E.~Woods, ``Software architecture in a changing world,'' \emph{IEEE Software},
  vol.~33, no.~6, pp. 94--97, Nov 2016.

\bibitem{Nazar2016}
\BIBentryALTinterwordspacing
N.~Nazar, Y.~Hu, and H.~Jiang, ``Summarizing software artifacts: A literature
  review,'' \emph{Journal of Computer Science and Technology}, vol.~31, no.~5,
  pp. 883--909, Sep 2016. [Online]. Available:
  \url{https://doi.org/10.1007/s11390-016-1671-1}
\BIBentrySTDinterwordspacing

\bibitem{Viviani2018}
G.~Viviani, C.~Janik-Jones, M.~Famelis, X.~Xia, and G.~C. Murphy, ``What design
  topics do developers discuss?'' in \emph{Proceedings of the IEEE
  International Conference on Program Comprehension}, 2018.

\bibitem{Viviani2018b}
G.~Viviani, C.~Janik-Jones, M.~Famelis, and G.~C. Murphy, ``The structure of
  software design discussions,'' in \emph{Proceedings of the International
  Workshop on Cooperative and Human Aspects of Software Engineering}.\hskip 1em
  plus 0.5em minus 0.4em\relax {ACM} Press, 2018.

\bibitem{Cubranic2003}
D.~{Cubranic} and G.~C. {Murphy}, ``Hipikat: recommending pertinent software
  development artifacts,'' in \emph{25th International Conference on Software
  Engineering, 2003. Proceedings.}, 2003, pp. 408--418.

\bibitem{Brunet2014}
J.~Brunet, G.~C. Murphy, R.~Terra, J.~Figueiredo, and D.~Serey, ``Do developers
  discuss design?'' in \emph{Working Conference on Mining Software
  Repositories}, Hyderabad, India, September 2014.

\bibitem{Shakiba2016}
\BIBentryALTinterwordspacing
A.~Shakiba, R.~Green, and R.~Dyer, ``{FourD}: do developers discuss design?
  revisited,'' in \emph{Proceedings of the 2nd International Workshop on
  Software Analytics - {SWAN} 2016}.\hskip 1em plus 0.5em minus 0.4em\relax
  {ACM} Press, 2016. [Online]. Available:
  \url{https://doi.org/10.1145/2989238.2989244}
\BIBentrySTDinterwordspacing

\bibitem{Viviani2019}
G.~{Viviani}, M.~{Famelis}, X.~{Xia}, C.~{Janik-Jones}, and G.~C. {Murphy},
  ``Locating latent design information in developer discussions: A study on
  pull requests,'' \emph{IEEE Transactions on Software Engineering}, pp. 1--1,
  2019.

\bibitem{bangash20}
A.~A. Bangash, H.~Sahar, A.~Hindle, and K.~Ali, ``On the time-based conclusion
  stability of cross-project defect prediction models,'' \emph{Empirical
  Software Engineering}, 2020.

\bibitem{igor2014}
\BIBentryALTinterwordspacing
I.~Steinmacher, I.~S. Wiese, T.~Conte, M.~A. Gerosa, and D.~Redmiles, ``The
  hard life of open source software project newcomers,'' in \emph{Proceedings
  of the 7th International Workshop on Cooperative and Human Aspects of
  Software Engineering}, ser. CHASE 2014.\hskip 1em plus 0.5em minus
  0.4em\relax New York, NY, USA: Association for Computing Machinery, 2014, p.
  72–78. [Online]. Available: \url{https://doi.org/10.1145/2593702.2593704}
\BIBentrySTDinterwordspacing

\bibitem{Bazelli2013}
B.~{Bazelli}, A.~{Hindle}, and E.~{Stroulia}, ``On the personality traits of
  stackoverflow users,'' in \emph{2013 IEEE International Conference on
  Software Maintenance}, 2013, pp. 460--463.

\bibitem{soliman2016}
M.~Soliman, M.~Galster, A.~R. Salama, and M.~Riebisch, ``Architectural
  knowledge for technology decisions in developer communities: An exploratory
  study with stackoverflow,'' in \emph{2016 13th Working IEEE/IFIP Conference
  on Software Architecture (WICSA)}.\hskip 1em plus 0.5em minus 0.4em\relax
  IEEE, 2016, pp. 128--133.

\bibitem{Mahadi_2020}
\BIBentryALTinterwordspacing
A.~Mahadi, K.~Tongay, and N.~A. Ernst, ``Cross-dataset design discussion
  mining,'' \emph{2020 IEEE 27th International Conference on Software Analysis,
  Evolution and Reengineering (SANER)}, Feb 2020. [Online]. Available:
  \url{http://dx.doi.org/10.1109/SANER48275.2020.9054792}
\BIBentrySTDinterwordspacing

\bibitem{Efstathiou2018}
\BIBentryALTinterwordspacing
V.~Efstathiou, C.~Chatzilenas, and D.~Spinellis, ``Word embeddings for the
  software engineering domain,'' in \emph{Proceedings of the 15th International
  Conference on Mining Software Repositories}, ser. MSR '18.\hskip 1em plus
  0.5em minus 0.4em\relax New York, NY, USA: Association for Computing
  Machinery, 2018, p. 38–41. [Online]. Available:
  \url{https://doi.org/10.1145/3196398.3196448}
\BIBentrySTDinterwordspacing

\bibitem{Zimmermann:2009}
T.~Zimmermann, N.~Nagappan, H.~Gall, E.~Giger, and B.~Murphy, ``Cross-project
  defect prediction: A large scale experiment on data vs. domain vs. process,''
  in \emph{Proceedings of the European Software Engineering Conference/ACM
  SIGSOFT International Symposium on Foundations of Software Engineering},
  2009, pp. 91--100.

\bibitem{herbold2017}
S.~Herbold, ``A systematic mapping study on cross-project defect prediction,''
  2017.

\bibitem{Pan:2010}
\BIBentryALTinterwordspacing
S.~J. Pan and Q.~Yang, ``A survey on transfer learning,'' \emph{IEEE
  Transactions on Knowledge and Data Engineering}, vol.~22, no.~10, pp.
  1345--1359, Oct. 2010. [Online]. Available:
  \url{https://doi.org/10.1109/TKDE.2009.191}
\BIBentrySTDinterwordspacing

\bibitem{Sharma:2019aa}
T.~Sharma, V.~Efstathiou, P.~Louridas, and D.~Spinellis, ``On the feasibility
  of transfer-learning code smells using deep learning,'' arXiv, Tech. Rep.
  1904.03031v2, 2019.

\bibitem{Krishna:2016:TMA:2970276.2970339}
\BIBentryALTinterwordspacing
R.~Krishna, T.~Menzies, and W.~Fu, ``Too much automation? the bellwether effect
  and its implications for transfer learning,'' in \emph{International
  Conference on Automated Software Engineering}, 2016, pp. 122--131. [Online].
  Available: \url{http://doi.acm.org/10.1145/2970276.2970339}
\BIBentrySTDinterwordspacing

\bibitem{hindlenatural}
A.~Hindle, E.~T. Barr, Z.~Su, M.~Gabel, and P.~Devanbu, ``On the naturalness of
  software,'' in \emph{Proceedings of the International Conference on Software
  Engineering}, 2012, p. 837–847.

\bibitem{Robbes:2019}
\BIBentryALTinterwordspacing
R.~Robbes and A.~Janes, ``Leveraging small software engineering data sets with
  pre-trained neural networks,'' in \emph{International Conference on Software
  Engineering: New Ideas and Emerging Results}, ser. ICSE-NIER '19, 2019, pp.
  29--32. [Online]. Available:
  \url{https://doi.org/10.1109/ICSE-NIER.2019.00016}
\BIBentrySTDinterwordspacing

\bibitem{Howard:2018}
J.~Howard and S.~Ruder, ``Universal language model fine-tuning for text
  classification,'' in \emph{Annual Meeting of the Association for
  Computational Linguistics}, 2018.

\bibitem{novielli20}
\BIBentryALTinterwordspacing
N.~Novielli, F.~Calefato, D.~Dongiovanni, D.~Girardi, and F.~Lanubile, ``Can we
  use se-specific sentiment analysis tools in a cross-platform setting?'' in
  \emph{International Conference on Mining Software Repositories}, 2020.
  [Online]. Available: \url{https://arxiv.org/abs/2004.00300}
\BIBentrySTDinterwordspacing

\bibitem{hindle11msr}
A.~Hindle, N.~Ernst, M.~W. Godfrey, and J.~Mylopoulos, ``{Automated topic
  naming to support cross-project analysis of software maintenance
  activities},'' in \emph{MSR}, Honolulu, 2011, pp. 1--10.

\bibitem{HindleBZN15}
\BIBentryALTinterwordspacing
A.~Hindle, C.~Bird, T.~Zimmermann, and N.~Nagappan, ``Do topics make sense to
  managers and developers?'' \emph{Empirical Software Engineering}, vol.~20,
  no.~2, pp. 479--515, 2015. [Online]. Available:
  \url{https://doi.org/10.1007/s10664-014-9312-1}
\BIBentrySTDinterwordspacing

\bibitem{Pickard1998}
\BIBentryALTinterwordspacing
L.~M. Pickard, B.~A. Kitchenham, and P.~W. Jones, ``Combining empirical results
  in software engineering,'' \emph{Information and Software Technology},
  vol.~40, no.~14, pp. 811 -- 821, 1998. [Online]. Available:
  \url{http://www.sciencedirect.com/science/article/pii/S0950584998001013}
\BIBentrySTDinterwordspacing

\bibitem{Aranda_2009}
\BIBentryALTinterwordspacing
J.~Aranda and G.~Venolia, ``The secret life of bugs: Going past the errors and
  omissions in software repositories,'' \emph{Proceedings of the ACM/IEEE
  International Conference on Software Engineering}, 2009. [Online]. Available:
  \url{http://dx.doi.org/10.1109/ICSE.2009.5070530}
\BIBentrySTDinterwordspacing

\bibitem{Ernst:2012wf}
N.~Ernst and G.~C. Murphy, ``{Case Studies in Just-In-Time Requirements
  Analysis},'' in \emph{Empirical Requirements Engineering Workshop at RE},
  Chicago, Sep. 2012, pp. 1--8.

\bibitem{Kitchenham2019}
\BIBentryALTinterwordspacing
B.~Kitchenham, L.~Madeyski, and P.~Brereton, ``Meta-analysis for families of
  experiments in software engineering: a systematic review and reproducibility
  and validity assessment,'' \emph{Empirical Software Engineering}, Jul 2019.
  [Online]. Available: \url{https://doi.org/10.1007/s10664-019-09747-0}
\BIBentrySTDinterwordspacing

\bibitem{Zanaty2018}
\BIBentryALTinterwordspacing
F.~E. Zanaty, T.~Hirao, S.~McIntosh, A.~Ihara, and K.~Matsumoto, ``An empirical
  study of design discussions in code review,'' in \emph{Proceedings of the
  International Symposium on Empirical Software Engineering and
  Measurement}.\hskip 1em plus 0.5em minus 0.4em\relax {ACM} Press, 2018.
  [Online]. Available: \url{https://doi.org/10.1145/3239235.3239525}
\BIBentrySTDinterwordspacing

\bibitem{Gelman:2013aa}
A.~Gelman and E.~Loken, ``The garden of forking paths: Why multiple comparisons
  can be a problem, even when there is no ``fishing expedition'' or
  ``p-hacking'' and the research hypothesis was posited ahead of time,''
  Colombia University, Tech. Rep., 2013.

\bibitem{Gelman:2012aa}
A.~Gelman, J.~Hill, and M.~Yajima, ``Why we (usually) don't have to worry about
  multiple comparisons,'' \emph{Journal of Research on Educational
  Effectiveness}, vol.~5, pp. 189--211, 2012.

\bibitem{DiNucci2018}
\BIBentryALTinterwordspacing
D.~D. Nucci, F.~Palomba, D.~A. Tamburri, A.~Serebrenik, and A.~D. Lucia,
  ``Detecting code smells using machine learning techniques: Are we there
  yet?'' in \emph{International Conference on Software Analysis, Evolution and
  Reengineering ({SANER})}, Mar. 2018. [Online]. Available:
  \url{https://doi.org/10.1109/saner.2018.8330266}
\BIBentrySTDinterwordspacing

\bibitem{Hill2012}
\BIBentryALTinterwordspacing
E.~Hill, S.~Rao, and A.~Kak, ``On the use of stemming for concern location and
  bug localization in java,'' in \emph{International Working Conference on
  Source Code Analysis and Manipulation}.\hskip 1em plus 0.5em minus
  0.4em\relax {IEEE}, Sep. 2012. [Online]. Available:
  \url{https://doi.org/10.1109/scam.2012.29}
\BIBentrySTDinterwordspacing

\bibitem{Menzies2012}
\BIBentryALTinterwordspacing
T.~Menzies and M.~Shepperd, ``Special issue on repeatable results in software
  engineering prediction,'' \emph{Empirical Software Engineering}, vol.~17,
  no.~1, pp. 1--17, Feb 2012. [Online]. Available:
  \url{https://doi.org/10.1007/s10664-011-9193-5}
\BIBentrySTDinterwordspacing

\bibitem{Kocaguneli2013}
\BIBentryALTinterwordspacing
E.~Kocaguneli and T.~Menzies, ``Software effort models should be assessed via
  leave-one-out validation,'' \emph{Journal of Systems and Software}, vol.~86,
  no.~7, pp. 1879--1890, jul 2013. [Online]. Available:
  \url{https://doi.org/10.1016/j.jss.2013.02.053}
\BIBentrySTDinterwordspacing

\bibitem{xia18}
\BIBentryALTinterwordspacing
T.~Xia, R.~Krishna, J.~Chen, G.~Mathew, X.~Shen, and T.~Menzies,
  ``Hyperparameter optimization for effort estimation,'' ArXiv, Tech. Rep.,
  2018. [Online]. Available: \url{http://arxiv.org/abs/1805.00336}
\BIBentrySTDinterwordspacing

\bibitem{Gmez2014}
\BIBentryALTinterwordspacing
O.~S. G{\'{o}}mez, N.~Juristo, and S.~Vegas, ``Understanding replication of
  experiments in software engineering: A classification,'' \emph{Information
  and Software Technology}, vol.~56, no.~8, pp. 1033--1048, aug 2014. [Online].
  Available: \url{https://doi.org/10.1016/j.infsof.2014.04.004}
\BIBentrySTDinterwordspacing

\bibitem{Storey:2019aa}
M.-A. Storey, C.~Williams, N.~A. Ernst, A.~Zagalsky, and E.~Kalliamvakou,
  ``Methodology matters: How we study socio-technical aspects in software
  engineering,'' arXiv, Tech. Rep. arXiv:1905.12841, 2019.

\bibitem{Shepperd2018}
M.~Shepperd, ``Replication studies considered harmful,'' in \emph{Companion of
  the International Conference on Software Engineering}, 2018.

\bibitem{mikolov2018advances}
T.~Mikolov, E.~Grave, P.~Bojanowski, C.~Puhrsch, and A.~Joulin, ``Advances in
  pre-training distributed word representations,'' in \emph{Proceedings of the
  International Conference on Language Resources and Evaluation (LREC 2018)},
  2018.

\bibitem{pennington2014glove}
\BIBentryALTinterwordspacing
J.~Pennington, R.~Socher, and C.~D. Manning, ``Glove: Global vectors for word
  representation,'' in \emph{Empirical Methods in Natural Language Processing
  (EMNLP)}, 2014, pp. 1532--1543. [Online]. Available:
  \url{http://www.aclweb.org/anthology/D14-1162}
\BIBentrySTDinterwordspacing

\bibitem{doc2vec14}
\BIBentryALTinterwordspacing
Q.~Le and T.~Mikolov, ``Distributed representations of sentences and
  documents,'' in \emph{Proceedings of the International Conference on Machine
  Learning}, 2014. [Online]. Available:
  \url{https://cs.stanford.edu/\~quocle/paragraph\_vector.pdf}
\BIBentrySTDinterwordspacing

\bibitem{BaltesDT008}
S.~Baltes, L.~Dumani, C.~Treude, and S.~Diehl, ``Sotorrent: reconstructing and
  analyzing the evolution of stack overflow posts,'' in \emph{Proceedings of
  the International Working Conference on Mining Software Repositories}, 2018,
  pp. 319--330.

\bibitem{Chawla:2002aa}
N.~Chawla, K.~Bowyer, L.~Hall, and W.~P. Kegelmeyer, ``Smote: Synthetic
  minority over-sampling technique,'' \emph{Journal of Artificial Intelligence
  Research}, vol.~16, pp. 321--357, 2002.

\bibitem{Alkadhi2017}
R.~Alkadhi, T.~Lata, E.~Guzmany, and B.~Bruegge, ``Rationale in development
  chat messages: An exploratory study,'' in \emph{Proceedings of the
  International Working Conference on Mining Software Repositories}, may 2017.

\bibitem{Maldonado2017}
\BIBentryALTinterwordspacing
E.~da~Silva~Maldonado, E.~Shihab, and N.~Tsantalis, ``Using natural language
  processing to automatically detect self-admitted technical debt,'' \emph{IEEE
  Transactions on Software Engineering}, vol.~43, no.~11, pp. 1044--1062, nov
  2017. [Online]. Available: \url{https://doi.org/10.1109/tse.2017.2654244}
\BIBentrySTDinterwordspacing

\bibitem{James2013}
\BIBentryALTinterwordspacing
G.~James, D.~Witten, T.~Hastie, and R.~Tibshirani, \emph{An Introduction to
  Statistical Learning: with Applications in R}.\hskip 1em plus 0.5em minus
  0.4em\relax Springer, 2013. [Online]. Available:
  \url{http://www-bcf.usc.edu/~gareth/ISL/getbook.html}
\BIBentrySTDinterwordspacing

\bibitem{karampatsis20}
\BIBentryALTinterwordspacing
R.-M. Karampatsis, H.~Babii, R.~Robbes, C.~Sutton, and A.~Janes, ``Big code !=
  big vocabulary,'' in \emph{Proceedings of the {ACM}/{IEEE} 42nd International
  Conference on Software Engineering}.\hskip 1em plus 0.5em minus 0.4em\relax
  {ACM}, Jun. 2020. [Online]. Available:
  \url{https://doi.org/10.1145/3377811.3380342}
\BIBentrySTDinterwordspacing

\bibitem{Wang2016}
\BIBentryALTinterwordspacing
P.~Wang, B.~Xu, J.~Xu, G.~Tian, C.-L. Liu, and H.~Hao, ``Semantic expansion
  using word embedding clustering and convolutional neural network for
  improving short text classification,'' \emph{Neurocomputing}, vol. 174, pp.
  806 -- 814, 2016. [Online]. Available:
  \url{http://www.sciencedirect.com/science/article/pii/S0925231215014502}
\BIBentrySTDinterwordspacing

\bibitem{li2015}
Y.~Li, L.~Xu, F.~Tian, L.~Jiang, X.~Zhong, and E.~Chen, ``Word embedding
  revisited: A new representation learning and explicit matrix factorization
  perspective,'' in \emph{Twenty-Fourth International Joint Conference on
  Artificial Intelligence}, 2015.

\bibitem{Jian2008}
\BIBentryALTinterwordspacing
J.~Hu, L.~Fang, Y.~Cao, H.-J. Zeng, H.~Li, Q.~Yang, and Z.~Chen, ``Enhancing
  text clustering by leveraging wikipedia semantics,'' in \emph{Proceedings of
  the 31st Annual International ACM SIGIR Conference on Research and
  Development in Information Retrieval}, ser. SIGIR ’08.\hskip 1em plus 0.5em
  minus 0.4em\relax New York, NY, USA: Association for Computing Machinery,
  2008, p. 179–186. [Online]. Available:
  \url{https://doi.org/10.1145/1390334.1390367}
\BIBentrySTDinterwordspacing

\bibitem{Sechidis}
K.~Sechidis, G.~Tsoumakas, and I.~Vlahavas, ``On the stratification of
  multi-label data,'' in \emph{Machine Learning and Knowledge Discovery in
  Databases}, D.~Gunopulos, T.~Hofmann, D.~Malerba, and M.~Vazirgiannis,
  Eds.\hskip 1em plus 0.5em minus 0.4em\relax Berlin, Heidelberg: Springer
  Berlin Heidelberg, 2011, pp. 145--158.

\bibitem{Hemalatha2012}
I.~Hemalatha, G.~S. Varma, and A.~Govardhan, ``Preprocessing the informal text
  for efficient sentiment analysis,'' \emph{International Journal of Emerging
  Trends \& Technology in Computer Science (IJETTCS)}, vol.~1, no.~2, pp.
  58--61, 2012.

\bibitem{Ghag2015}
K.~V. {Ghag} and K.~{Shah}, ``Comparative analysis of effect of stopwords
  removal on sentiment classification,'' in \emph{2015 International Conference
  on Computer, Communication and Control (IC4)}, 2015, pp. 1--6.

\bibitem{Balakrishnan2014}
V.~Balakrishnan and L.-Y. Ethel, ``Stemming and lemmatization: A comparison of
  retrieval performances,'' \emph{Lecture Notes on Software Engineering},
  vol.~2, no.~3, p. 262–267, 2014.

\bibitem{Miller1998}
G.~A. Miller, \emph{WordNet: An electronic lexical database}.\hskip 1em plus
  0.5em minus 0.4em\relax MIT press, 1998.

\bibitem{tremblay2011}
A.~Tremblay and B.~V. Tucker, ``The effects of n-gram probabilistic measures on
  the recognition and production of four-word sequences,'' \emph{The Mental
  Lexicon}, vol.~6, no.~2, pp. 302--324, 2011.

\bibitem{joulin2016}
A.~Joulin, E.~Grave, P.~Bojanowski, and T.~Mikolov, ``Bag of tricks for
  efficient text classification,'' \emph{arXiv preprint arXiv:1607.01759},
  2016.

\bibitem{joulin2016a}
A.~Joulin, E.~Grave, P.~Bojanowski, M.~Douze, H.~J{\'e}gou, and T.~Mikolov,
  ``Fasttext.zip: Compressing text classification models,'' \emph{arXiv
  preprint arXiv:1612.03651}, 2016.

\bibitem{bojanowski2016}
P.~Bojanowski, E.~Grave, A.~Joulin, and T.~Mikolov, ``Enriching word vectors
  with subword information,'' \emph{arXiv preprint arXiv:1607.04606}, 2016.

\bibitem{mikolov2013}
T.~Mikolov, K.~Chen, G.~Corrado, and J.~Dean, ``Efficient estimation of word
  representations in vector space,'' 2013.

\bibitem{mikolov2013skipgram_improve}
T.~Mikolov, I.~Sutskever, K.~Chen, G.~Corrado, and J.~Dean, ``Distributed
  representations of words and phrases and their compositionality,'' 2013.

\bibitem{scikit-learn}
F.~Pedregosa, G.~Varoquaux, A.~Gramfort, V.~Michel, B.~Thirion, O.~Grisel,
  M.~Blondel, P.~Prettenhofer, R.~Weiss, V.~Dubourg, J.~Vanderplas, A.~Passos,
  D.~Cournapeau, M.~Brucher, M.~Perrot, and E.~Duchesnay, ``Scikit-learn:
  Machine learning in {P}ython,'' \emph{Journal of Machine Learning Research},
  vol.~12, pp. 2825--2830, 2011.

\bibitem{Menzies2019}
\BIBentryALTinterwordspacing
T.~Menzies and M.~Shepperd, ````{B}ad smells'' in software analytics papers,''
  \emph{Information and Software Technology}, vol. 112, pp. 35--47, Aug. 2019.
  [Online]. Available: \url{https://doi.org/10.1016/j.infsof.2019.04.005}
\BIBentrySTDinterwordspacing

\bibitem{Sun2017}
\BIBentryALTinterwordspacing
C.~Sun, A.~Shrivastava, S.~Singh, and A.~Gupta, ``Revisiting unreasonable
  effectiveness of data in deep learning era,'' in \emph{2017 {IEEE}
  International Conference on Computer Vision ({ICCV})}.\hskip 1em plus 0.5em
  minus 0.4em\relax {IEEE}, Oct. 2017. [Online]. Available:
  \url{https://doi.org/10.1109/iccv.2017.97}
\BIBentrySTDinterwordspacing

\bibitem{romano20}
\BIBentryALTinterwordspacing
S.~Romano, D.~Fucci, G.~Scanniello, M.~T. Baldasarre, B.~Turhan, and
  N.~Juristo, ``Researcher bias in software engineering experiments: a
  qualitative investigation,'' in \emph{Software Engineering and Advanced
  Applications}, 2020. [Online]. Available:
  \url{https://arxiv.org/abs/2008.12528v1}
\BIBentrySTDinterwordspacing

\bibitem{vanVliet2016}
H.~van Vliet and A.~Tang, ``Decision making in software architecture,''
  \emph{Journal of Systems and Software}, vol. 117, pp. 638--644, jul 2016.

\bibitem{Alkadhi2018}
\BIBentryALTinterwordspacing
R.~Alkadhi, M.~Nonnenmacher, E.~Guzman, and B.~Bruegge, ``How do developers
  discuss rationale?'' in \emph{Proceedings of the IEEE International
  Conference on Software Analysis, Evolution, and Reengineering}.\hskip 1em
  plus 0.5em minus 0.4em\relax {IEEE}, mar 2018. [Online]. Available:
  \url{https://doi.org/10.1109/saner.2018.8330223}
\BIBentrySTDinterwordspacing

\bibitem{Motta2018}
\BIBentryALTinterwordspacing
T.~O. Motta, R.~R. {Gomes e Souza}, and C.~Sant'Anna, ``Characterizing
  architectural information in commit messages,'' in \emph{Proceedings of the
  Brazilian Symposium on Software Engineering}.\hskip 1em plus 0.5em minus
  0.4em\relax {ACM} Press, 2018. [Online]. Available:
  \url{https://doi.org/10.1145/3266237.3266260}
\BIBentrySTDinterwordspacing

\bibitem{Arya2019}
\BIBentryALTinterwordspacing
D.~Arya, W.~Wang, J.~L. Guo, and J.~Cheng, ``Analysis and detection of
  information types of open source software issue discussions,'' in
  \emph{International Conference on Software Engineering ({ICSE})}.\hskip 1em
  plus 0.5em minus 0.4em\relax {IEEE}, May 2019. [Online]. Available:
  \url{https://doi.org/10.1109/icse.2019.00058}
\BIBentrySTDinterwordspacing

\end{thebibliography}
